\documentclass{aa}  

\usepackage{natbib}
\usepackage{graphicx}
\usepackage{txfonts}
\usepackage{hyperref}
\usepackage{color}
\usepackage{upgreek}  
\usepackage{color}
\usepackage[version=3]{mhchem}
\begin{document} 
   \title{Carbon star wind models at solar and sub-solar metallicities: \\ a comparative study}
   \subtitle{I. Mass loss and the properties of dust-driven winds}

   \author{S. Bladh\inst{1,2}
          \and
          K. Eriksson\inst{2}
          \and 
          P. Marigo\inst{1}
          \and
          S. Liljegren\inst{2}
          \and
		  B. Aringer\inst{1}
                    }

   \institute{Dipartimento di Fisica e Astronomia Galileo Galilei, Università di Padova, Vicolo dell’Osservatorio 3, 35122 Padova, Italy\label{inst2}\\
         	\and
            Theoretical Astrophysics, Department of Physics and Astronomy, Uppsala University, Box 516, SE-751 20 Uppsala, Sweden\label{inst1}
            \\
            \email{sara.bladh@physics.uu.se}
             }

   \date{Received 04/12/2018; accepted 28/01/2019}
   
 
  \abstract
  {The heavy mass loss observed in evolved stars on the asymptotic giant branch (AGB) is usually attributed to dust-driven winds, but it is still an open question how much AGB stars contribute to the dust production in the interstellar medium, especially at lower metallicities. In the case of C-type AGB stars, where the wind is thought to be driven by radiation pressure on amorphous carbon grains, there should be significant dust production even in metal-poor environments. Carbon stars can manufacture the building blocks needed to form the wind-driving dust species themselves, irrespective of the chemical composition they have, by dredging up carbon from the stellar interior during thermal pulses.}
   {We investigate how the mass loss in carbon stars is affected by a low-metallicity environment, similar to the Large and Small Magellanic Clouds (LMC and SMC).}
   {The atmospheres and winds of C-type AGB stars are modeled with the 1D spherically symmetric radiation-hydrodynamical code Dynamic Atmosphere and Radiation-driven Wind models based on Implicit Numerics (DARWIN). The models include a time-dependent description for nucleation, growth, and evaporation of amorphous carbon grains directly out of the gas phase. To explore the metallicity-dependence of mass loss we calculate model grids at three different chemical abundances (solar, LMC, and SMC).  Since carbon may be dredged up during the thermal pulses as AGB stars evolve, we keep the carbon abundance as a free parameter. The models in these three different grids all have a current mass of one solar mass; effective temperatures of 2600\,K, 2800\,K, 3000\,K, or 3200\,K; and stellar luminosities equal to $\log L_*/L_{\odot}=3.70$, 3.85, or 4.00.}
   {The DARWIN models show that mass loss in carbon stars is facilitated by high luminosities, low effective temperatures, and a high carbon excess (C-O) at both solar and subsolar metallicities.  Similar combinations of effective temperature, luminosity, and carbon excess produce outflows at both solar and subsolar metallicities. There are no large systematic differences in the mass-loss rates and wind velocities produced by these wind models with respect to metallicity, nor any systematic difference concerning the distribution of grain sizes or how much carbon is condensed into dust. DARWIN models at subsolar metallicity have approximately 15\% lower mass-loss rates compared to DARWIN models at solar metallicity with the same stellar parameters and carbon excess. For both solar and subsolar environments typical grain sizes range between $0.1$ and $0.5\,\mu$m, the degree of condensed carbon varies between 5\% and 40\%, and the gas-to-dust ratios between 500 and 10000.}
   {C-type AGB stars can contribute to the dust production at subsolar metallicities (down to at least $\mathrm{[Fe/H]}=-1$) as long as they dredge up sufficient amounts of carbon from the stellar interior. Furthermore, stellar evolution models can use the mass-loss rates calculated from DARWIN models at solar metallicity when modeling the AGB phase at subsolar metallicities if carbon excess is used as the critical abundance parameter instead of the C/O ratio.}
   {}


   \maketitle

\section{Introduction}
\label{s_intro}
Mass loss plays an important role in stellar evolution modeling during the asymptotic giant branch (AGB) phase, for estimating the lifetime of individual AGB stars and evaluating how much AGB stars contribute to the dust production and enrichment of heavier elements in the interstellar medium. Even a modest change in the mass-loss rate, e.g., a factor of two, will have a strong effect on the evolution of an AGB star and its mass return to the interstellar medium \citep[e.g.,][]{Lattanzio2016}. Since mass loss, through stellar winds, determines the duration of the AGB phase it also influences the mass range of white dwarfs \citep[e.g.,][]{Kalirai_etal_14} and the contribution of AGB stars to the integrated light of galaxies \citep[e.g.,][]{marigo2017, Marigo_15}.
Wind models that can predict accurate mass-loss rates are therefore crucial for stellar evolution models focusing on the AGB phase. This is especially important at lower metallicities where it is still an open question how much AGB stars contribute to the interstellar dust production and the cosmic matter cycle. 

In M-type AGB stars the stellar wind is probably driven by photon scattering on Mg/Fe-silicates \citep{hoefner08,bladh2012,bladh2013,bladh2015,hoefner2016}. It should therefore be more difficult to produce outflows in M-type AGB stars, i.e., require more extreme stellar parameters, in metal-poor environments since the necessary elements are less abundant, which results in less dust material to drive a wind. In the case of C-type AGB stars, where the wind is thought to be driven by radiation pressure on amorphous carbon grains \citep[e.g.][]{winters2000,wachter2002,wachter2008,hoefner2003,mattsson2010,eriksson14}, the situation is different. Carbon stars can manufacture the constituents needed to form the wind-driving dust species themselves, irrespective of the chemical composition they have, by dredging up carbon from the interior during the thermal pulses. There should therefore be significant dust production even in metal-poor environments, especially since the number of carbon stars per unit mass seem to increase with decreasing metallicity, at least in the Local Group \citep[e.g.][]{westerlund1983,rossi1999,boyer2013,cioni2003}. This increased frequency of carbon stars is a consequence of the lower oxygen abundance in metal-poor environments, since less carbon has to be dredged up into the atmosphere to make C/O>1.

Observationally, the mass-loss rates for carbon stars can be estimated either from the emission observed in circumstellar CO rotational lines or from the observed dust emission \citep[see Sect.~2.3 in][for further details]{hoefner2018}. The method using CO-line emission is generally considered more reliable since fewer assumptions are made about properties that cannot be measured directly (e.g., the dust velocity). However, it is limited in its reach. For large samples of distant sources, such as AGB stars in the Magellanic Clouds, mass-loss rate estimates based on dust emission is essentially the only method available. This introduces a problem when investigating the metallicity dependence of mass loss in AGB stars since it is not straightforward to compare wind properties derived from these two different methods, unless care is taken to calculate them consistently. Another complication is the uncertainty when determining distances to Galactic AGB stars, which prevents accurate estimates of luminosity and mass-loss rate. Observing sources at known distances, such as the Magellanic Clouds, reduces this problem.

Comparisons between observational studies of carbon stars in the Magellanic Clouds and the Milky Way, based on dust-emission estimates, indicate that mass-loss rates in solar and subsolar environments are similar \citep[e.g.][]{vanloon2000,Groenewegen2007}. It should be noted that that in \cite{Groenewegen2016} the wind velocity is assumed to be similar to that observed for Galactic carbon stars, whereas in \cite{vanloon2000} the wind velocity was scaled with metallicity. \cite{Groenewegen2016} recently published wind properties for four extreme carbon stars in the Large Magellanic Clouds, estimated for the first time from observations of rotational CO-lines by ALMA. For this observational sample, although it is small, the correlation between wind properties differs from the observed in samples of nearby Galactic carbon stars. Noticeably, in this sample the wind velocity is generally lower than that seen in nearby Galactic carbon stars. 

To help understand these observational findings and to investigate how mass loss in carbon stars is affected by a metal-poor composition, we present here the results from dynamical models of the atmospheres and winds of C-type AGB stars, using the 1D radiation-hydrodynamic code Dynamic Atmosphere and Radiation-driven Wind models based on
Implicit Numerics (DARWIN). The metallicity dependence is explored by calculating model grids at three different metallicities, corresponding to a solar-like environment and subsolar environments similar to the Large and Small Magellanic Clouds (LMC and SMC). A first exploratory work using DARWIN models of C-type AGB stars at subsolar metallicities was presented by \cite{mattsson2008}. In that paper, dynamical models with two sets of stellar parameters, as well as chemical compositions corresponding to solar and LMC metallicity, were explored. The resulting mass-loss rates from wind models with different chemical composition were never directly compared, but it was concluded that it is unlikely that metal-poor carbon stars have lower mass-loss rates than their more metal-rich counterparts with similar stellar parameters, as long as they have a comparable amount of condensible carbon. The aim of this study is to produce grids of DARWIN models at solar and subsolar metallicities, covering a larger sample of stellar parameters, and then directly compare how wind and dust properties are affected by a decrease in metallicity.

This paper is organized in the following way. In Sect.~\ref{s_darwin} we summarize the main features of the DARWIN code for C-type AGB stars. In Sect.~\ref{s_gridpar} we present the input parameters of the model grids, and details concerning the spectral synthesis is given in Sect.~\ref{s_sed}. The resulting wind and grain properties, and how they depend on different input parameters and metallicity, are presented in Sect.~\ref{s_dynres}. In Sect.~\ref{s_mol} we evaluate how changes in the molecular abundances influence the radiation field in subsolar environments. In Sect.\ref{s_sem} we discuss the implication for stellar evolution models. Finally, in Sect.~\ref{s_concl} we give a short summary of our conclusions.

\begin{figure}
\centering
\includegraphics[width=\linewidth]{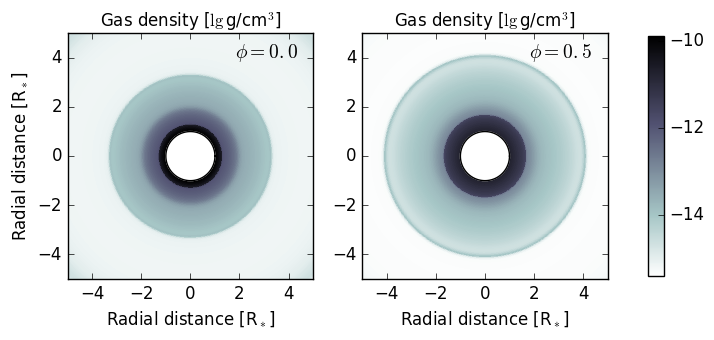}\\
\includegraphics[width=\linewidth]{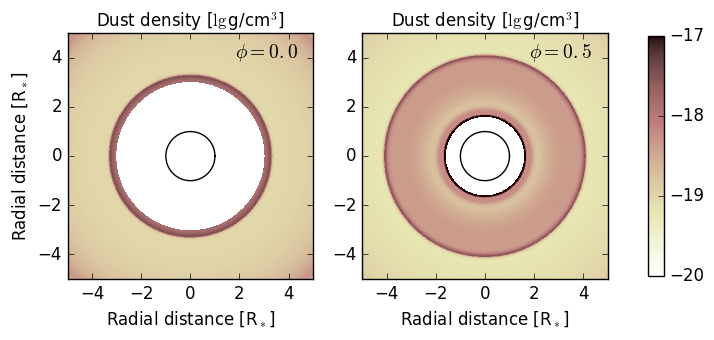}\\
\includegraphics[width=\linewidth]{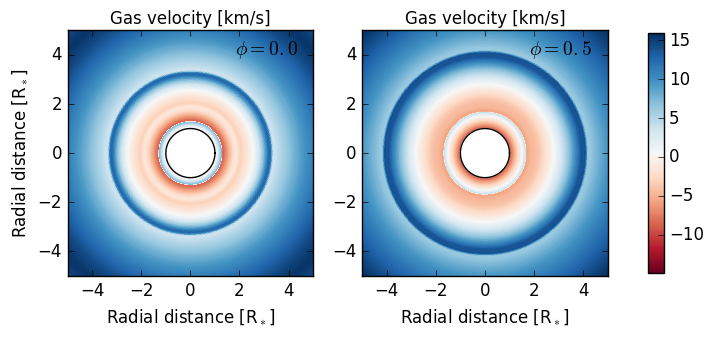}
   \caption{Cross-sectional snapshots of a DARWIN model for a C-type AGB star, with $M_*=1\,\mathrm{M}_{\odot}$, $\log L_*/L_{\odot}=4.00$, $T_*=2600\,$K, $u_{\mathrm{p}}=6$\,km/s, and $[\mathrm{Fe/H}]=-1.0$ (SMC). The upper, middle, and lower panels show gas density, dust density, and velocity of the gas (negative velocity inward), respectively. The panels show atmospheric snapshots at maximum (left column, $\phi=0.0$) and minimum (right column, $\phi=0.5$) luminosity. The solid black line indicates the radius of the star, calculated from the Stefan-Boltzmann law.}
    \label{f_structure}
\end{figure}

\section{Modeling methods: dynamical models}
\label{s_darwin}
Stellar winds in evolved AGB stars are generally considered to be pulsation-enhanced
dust-driven outflows \citep[PEDDRO scenario, see][]{hoefner2018}.
This mass-loss scenario is built on a two-stage process, where in the first stage the contracting and expanding photosphere of the star gives rise to sound waves. These sound waves develop into shock waves when they propagate through the steep density gradient of the atmosphere, thereby creating favorable conditions for grain formation. In the second stage the newly formed dust particles are accelerated away from the star by momentum transfer from stellar photons, either through photon scattering or photon absorption. Friction between the accelerated dust particles and the surrounding gas then triggers a general outflow. For carbon stars these outflows are mainly driven by photon absorption on amorphous carbon grains. Detailed models of this scenario show good agreement with a range of observations, for example high resolution spectroscopy, photometry, and interferometry \citep[e.g.][]{winters2000,Gautschy2004,nowotny10,nowotny11,sacuto11,eriksson14,liljegren16,rau17,witt17}.

\begin{table*}
\caption{The chemical abundances, the carbon excess and the C/O ratio for the model grids presented in this study.}   
\label{t_opacity}      
\centering          
\begin{tabular}{l r c c c c c}   
\hline\hline     
Metallicity  & [Fe/H] & log[C]+12 & log[O]+12 & log[N]+12 & log[C-O]+12 & C/O \\ 
environments &        &    [dex]  & [dex]     & [dex]     & [dex]       &     \\
\hline
Solar &  $0.0$ & 8.79 & 8.66 & 7.78 & 8.20 & 1.35 \\
Solar &  $0.0$ & 8.89 & 8.66 & 7.78 & 8.50 & 1.69 \\
Solar &  $0.0$ & 9.04 & 8.66 & 7.78 & 8.80 & 2.38 \\
\hline
LMC   & $-0.5$ & 8.48 & 8.16 & 7.28 & 8.20 & 2.10 \\
LMC   & $-0.5$ & 8.66 & 8.16 & 7.28 & 8.50 & 3.19 \\
LMC   & $-0.5$ & 8.89 & 8.16 & 7.28 & 8.80 & 5.37 \\
\hline
SMC   & $-1.0$ & 8.31 & 7.66 & 6.78 & 8.20 & 4.47  \\
SMC   & $-1.0$ & 8.56 & 7.66 & 6.78 & 8.50 & 7.92  \\
SMC   & $-1.0$ & 8.83 & 7.66 & 6.78 & 8.80 & 14.80 \\
\hline
\end{tabular}
\end{table*}

\begin{table}
\caption{The input parameters covered by the model grids.}   
\label{t_grid}      
\centering          
\begin{tabular}{l l l | l l l | l}   
\hline\hline     
$M_*$           & $T_*$ & $\log L_*/L_{\odot}$  & $P$ & $\Delta u_{\mathrm{p}}$ & $f_{\mathrm{L}}$ & $\log[\mathrm{C-O}]$ \\ 
~[$M_{\odot}$] & [K]   & & [d] & [km/s]                  &                     & +12 [dex]          \\
\hline
1.0 & 2600 & 3.70 & 294 & 2, 4, 6 & 2 & 8.2, 8.5, 8.8\\
    &      & 3.85 & 390 & 2, 4, 6 & 2 & 8.2, 8.5, 8.8\\
    &      & 4.00 & 525 & 2, 4, 6 & 2 & 8.2, 8.5, 8.8\\
\hline
1.0 & 2800 & 3.70 & 294 & 2, 4, 6 & 2 & 8.2, 8.5, 8.8\\
    &      & 3.85 & 390 & 2, 4, 6 & 2 & 8.2, 8.5, 8.8\\
    &      & 4.00 & 525 & 2, 4, 6 & 2 & 8.2, 8.5, 8.8\\
\hline
1.0 & 3000 & 3.70 & 294 & 2, 4, 6 & 2 & 8.2, 8.5, 8.8\\
    &      & 3.85 & 390 & 2, 4, 6 & 2 & 8.2, 8.5, 8.8\\
    &      & 4.00 & 525 & 2, 4, 6 & 2 & 8.2, 8.5, 8.8\\
\hline
1.0 & 3200 & 3.70 & 294 & 2, 4, 6 & 2 & 8.2, 8.5, 8.8\\
    &      & 3.85 & 390 & 2, 4, 6 & 2 & 8.2, 8.5, 8.8\\
    &      & 4.00 & 525 & 2, 4, 6 & 2 & 8.2, 8.5, 8.8\\    
\hline    
\end{tabular}
\tablefoot{The stellar parameters (mass, effective temperature and luminosity) are listed in columns 1-3. The pulsation parameters (period, piston velocity and scaling factor for the luminosity amplitude) describing the sinusoidal variations at the inner boundary are listed in columns 3-6. Lastly, the carbon excess is listed in column 7. Note that the period $P$ is not an independent parameter, but coupled to the
stellar luminosity by the period-luminosity relation in \cite{feast1989}.}
\end{table}

\subsection{DARWIN models}

The atmospheres and winds of C-type AGB stars are modeled using the 1D spherically symmetric radiation-hydrodynamical code DARWIN (Dynamic Atmosphere and Radiation-driven Wind models based on Implicit Numerics). For a detailed description of the modeling methods and equations included in the DARWIN code, see \cite{hoefner2003,hoefner2016}. The DARWIN models cover a spherical shell, with an inner boundary situated just below the stellar photosphere but above the driving zone of the pulsation.  
The location of the outer boundary is determined by the dynamics of the model atmosphere. If an outflow develops the outer boundary is set at 25 stellar radii, where the flow velocities typically have reached terminal value. Otherwise, the outer boundary follows the periodic motion of the upper atmospheric layers.

The initial atmospheric structure of a DARWIN model is a classic hydrostatic model atmosphere, characterized by the fundamental stellar parameters (mass $M_*$, luminosity $L_*$, effective temperature $T_*$) and the chemical composition. The effects of stellar pulsations are simulated by sinusoidal variations of radius and luminosity imposed on the innermost mass shell. The amplitude of this variation is gradually ramped up over about 50 pulsation cycles, turning the initially hydrostatic atmosphere into a dynamical atmosphere. For each time-step the dynamics of the system is described by the equations of hydrodynamics (conservation of mass, momentum, and energy), the frequency-integrated zeroth and first moment equations of the radiative transfer equation (accounting for the energy and momentum balance of the radiative field), and the moment equations describing time-dependent growth of dust grains. We assume complete momentum coupling of gas and dust, i.e., momentum gained by the dust from the radiation field is distributed to the gas by collisions and they are dynamically treated as one fluid. This system of partial differential equations is solved with a Newton-Raphson scheme on an adaptive spatial grid \citep{dorfi1987}. The calculations presented here use 100 radial grid points and the resolution of the adaptive grid is determined by taking temperature and density gradients into account.

The gas opacities are calculated with the opacity generation program COMA \citep{aringer2016} and include 25 different molecular species. The chemical abundances used for the gas opacity and dust calculations are consistent in COMA and DARWIN. The radiative transfer equation is solved at 64 frequency points. These frequency points are roughly equidistant in wavenumber (and therefore in energy), but are randomly distributed relative to opacity features, as required by the opacity sampling approach. Since the opacities of the molecules and dust forming in the extended atmosphere dominate the radiation field, the inclusion of frequency-dependent radiative transfer is crucial for achieving realistic atmospheric structures \citep{hoefner2003}.

Figure~\ref{f_structure} shows the gas density (upper panels), dust density (middle panels), and velocity (lower panels) in the atmospheric structures at maximum and minimum luminosity (at $\phi=0.0$ and $\phi=0.5$, respectively) taken from a DARWIN model. As can be seen, the gas layers move inward and outward in the inner atmosphere, before the dust condenses and the material is accelerated away by radiation pressure on dust.

\begin{figure}
\centering
\includegraphics[width=\linewidth]{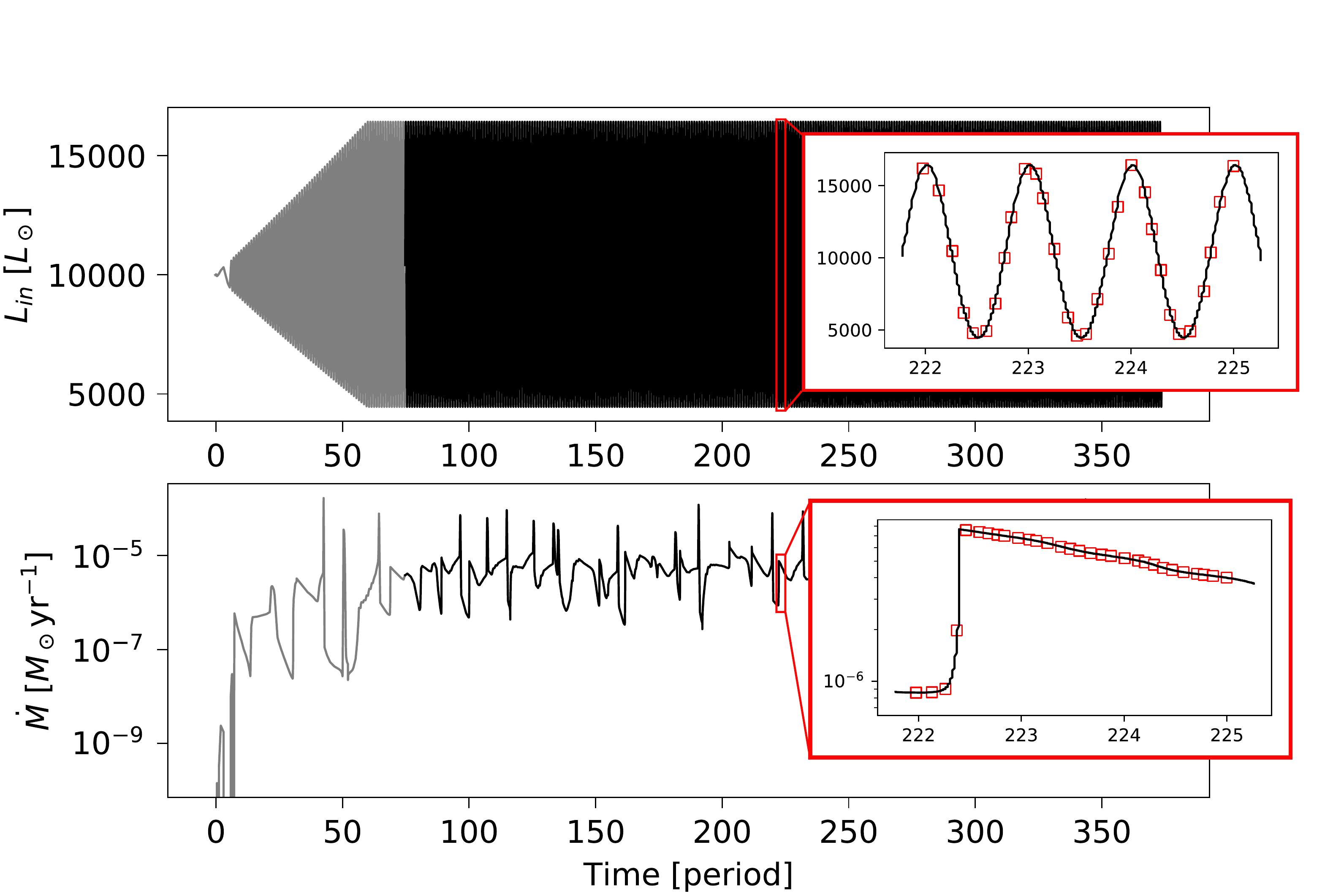}
   \caption{Temporal evolution of a DARWIN model with input parameters $M_*=1\,\mathrm{M}_{\odot}$, $\log L_*/L_{\odot}=4.00$, $T_*=2600\,$K, $u_{\mathrm{p}}=6$\,km/s, $\log(\mathrm{C-O})+12=8.5$, and $[\mathrm{Fe/H}]=-1.0$ (SMC). The panels show luminosity and the resulting mass-loss rate at the outer boundary. The wind properties for this model are calculated by averaging over the time interval marked in black. The two embedded panels show the snapshots selected for the a posteriori radiative transfer in red.}
    \label{f_timeseries}
\end{figure}

\subsection{Pulsation description}
\label{s_puls}
The stellar pulsations are described by sinusoidal variations of radius and luminosity at the inner boundary. In this "piston model" scheme \citep{bowen88}, the radial variation at the inner boundary $R_{in}(t)$ is given by 

\begin{equation}
\label{eqn1}
R_\mathrm{in}(t) = R_0 + \frac{\Delta u_p P}{2 \pi} \sin{\left ( \frac{2 \pi }{P} t \right)},
\end{equation}
where $\Delta u_p$ is the velocity amplitude and $P$ is the pulsation period. The corresponding velocity variation is given by
\begin{equation}
\label{eqn11}
u_\mathrm{in}(t) = \Delta u_p \cos\left ( \frac{2 \pi }{P} t \right).
\end{equation}
The luminosity variation at the inner boundary $L_\mathrm{in}(t)$ is proportional to the square of the radial variation and has the same periodicity. Moreover, the luminosity amplitude can be adjusted separately with the scaling factor $f_{\mathrm{L}}$, 
\begin{equation}
\label{eqn2}
\Delta L_\mathrm{in}(t) = L_\mathrm{in} - L_0= f_L \left (\frac{R^2_{in}(t) - R^2_0}{R^2_0} \right ) \times L_0.
\end{equation}
It is important to keep in mind that the inner boundary condition used here describes pulsation properties (period and amplitude) of Mira variables pulsating in the fundamental mode. For a more in depth discussion concerning the pulsation description and its effect on the models, see \cite{liljegren16}.

\subsection{Dust description}
The wind-driving dust species in DARWIN models for C-type AGB stars is amorphous carbon, a carbon compound with a mixture of $sp^2$ and $sp^3$ bonds. The DARWIN models include a description for the nucleation of amorphous carbon seed particles, based on classical nucleation theory. The growth and evaporation of the amorphous carbon grains are assumed to proceed according to net reactions\\
\newline
\indent\ce{C2H2 + C_N <-> C_{N+2} + H2}\\
\newline
\indent\ce{C2H + C_N <-> C_{N+2} + H}\\
\newline
involving the species C, C$_2$, C$_2$H, and C$_2$H$_2$, accounting for chemical sputtering \citep[see][]{hoefner1995,gauger90}. Strong molecular bands of C$_2$H$_2$ have been observed in Galactic carbon stars, but also in carbon stars at subsolar metallicities such as the Magellanic Clouds \citep[e.g.][]{loon1999,matsuura2005,loon2006,sloan2006,zijlstra2006,matsuura2006,lagadec2007,loon2008}, suggesting that these reactions for grain formation are relevant even in metal-poor environments. The abundance of these molecular species is determined by chemical equilibrium and the material condensed into dust particles is subtracted from the gas phase at each time-step. 

The time-dependent description of grain growth and evaporation follows the moment method developed by \cite{gail88} and \cite{gauger90}. In this approach, the dust particles at distance $r$ from the stellar center and at time $t$ are described in terms of moments $K_{i}(r, t)$ of the grain size distribution function $n(a_{\mathrm{gr}},r, t)$, weighted with a power $i$ of the grain radius $a_{\mathrm{gr}}$,
\begin{equation*}
K_i(r,t) \propto \int_0^{\infty}a_{\mathrm{gr}}^{i}\,n(a_{\mathrm{gr}}(r, t))\,\mathrm{d}a_{\mathrm{gr}}\qquad (i=0,1,2,3).
\end{equation*}
It follows from this definition that $K_0$ is proportional to the total number density of grains, while $K_1$, $K_2$, and $K_3$ are related to the average radius, geometric cross section, and volume of the grains, respectively.
Solving the differential equations that determine the evolution of the dust moments $K_{i}$ gives information about the particle radii in every layer at every instance of time. The average grain radius, calculated from the moment $K_1$ together with the moment $K_3$, are used to compute size-dependent dust opacities and determining the radiative acceleration of the dust-gas mixture. The dust grains are assumed to be spherical and the optical properties are calculated from data by \cite{rolmar91} using Mie theory. For a detailed description of the treatment of size-dependent opacities using the moment method, see \cite{mattsson2011}.

\begin{figure}
\centering
\includegraphics[width=0.8\linewidth]{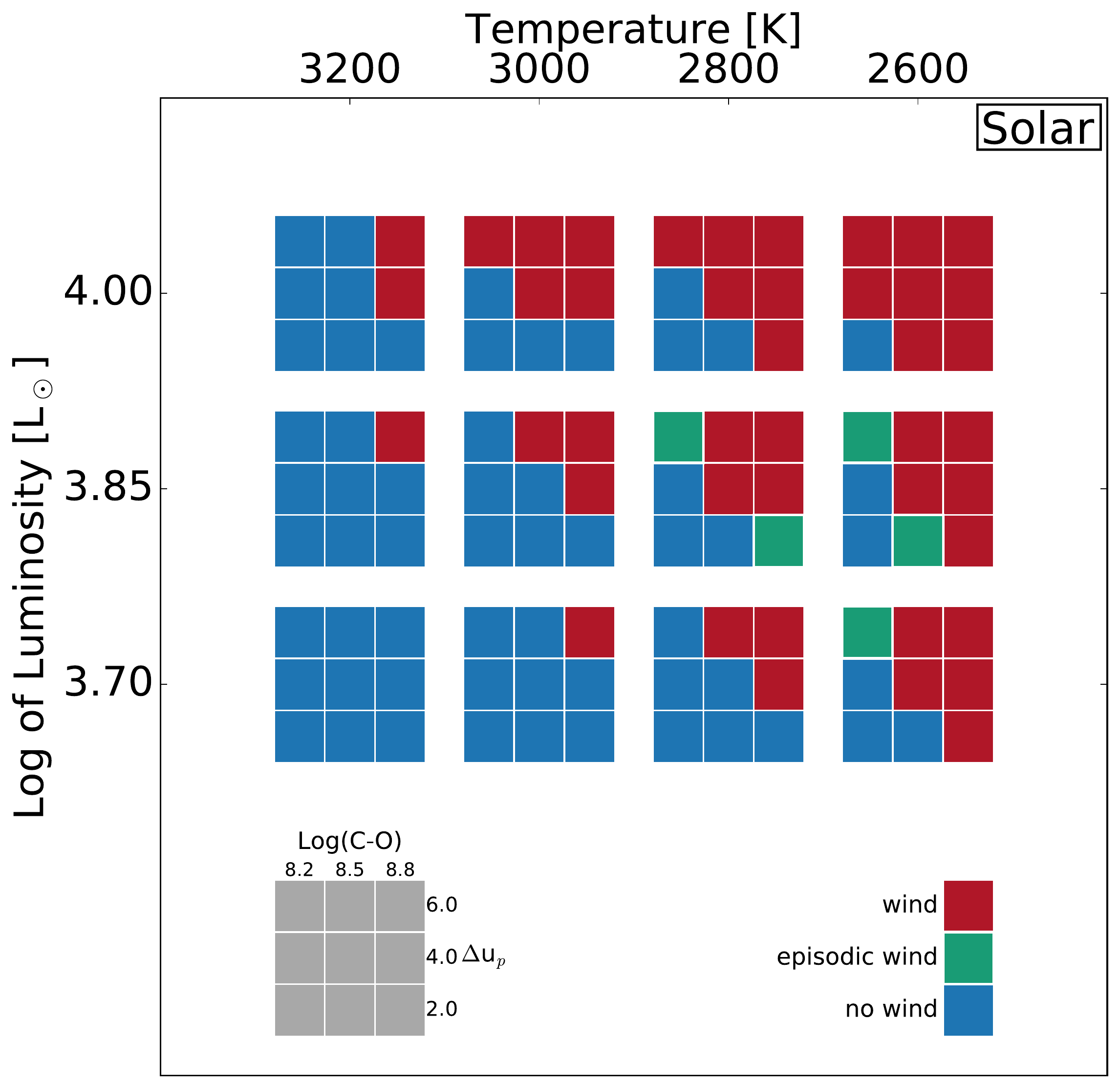}\\
\includegraphics[width=0.8\linewidth]{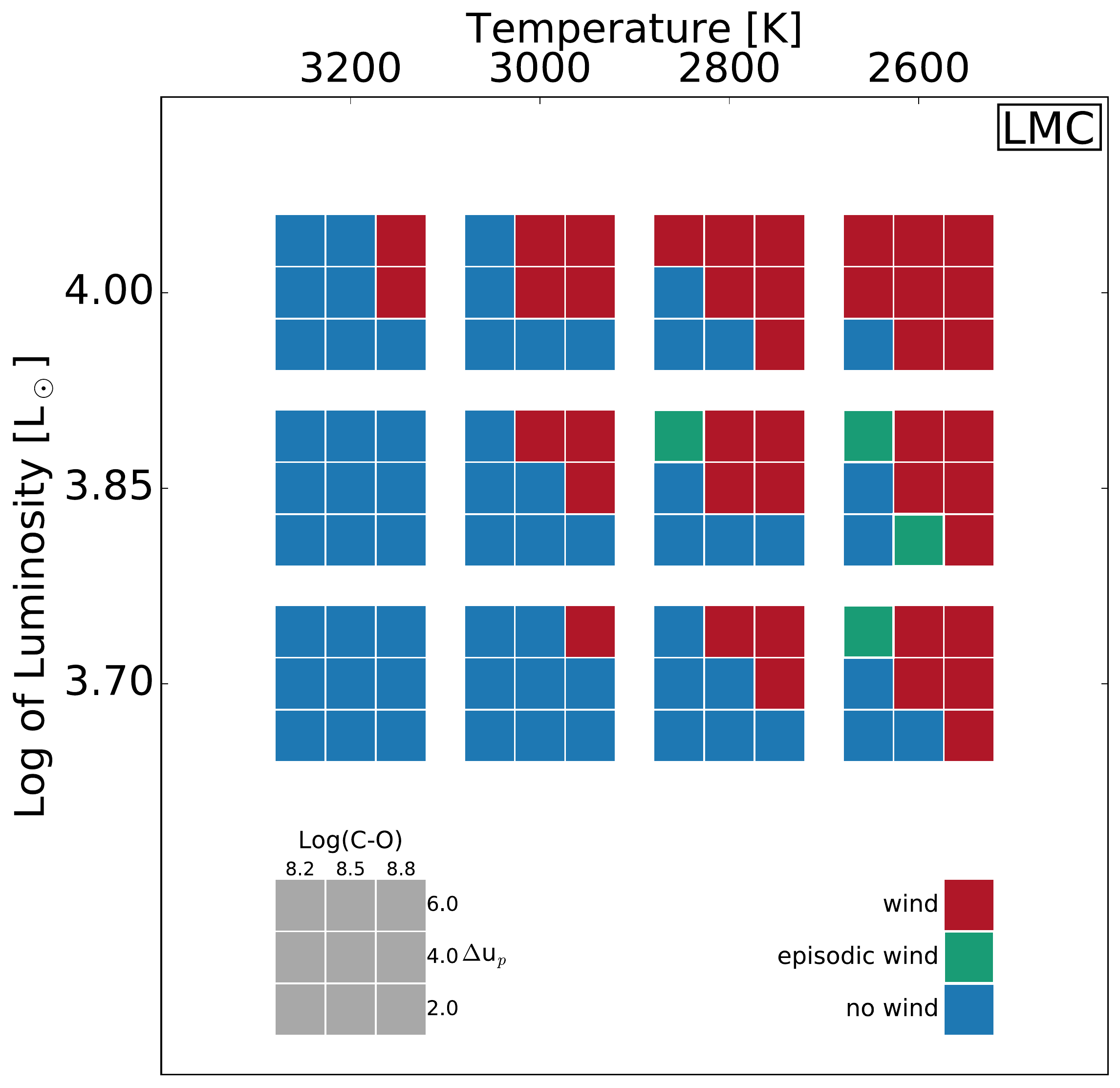}\\
\includegraphics[width=0.8\linewidth]{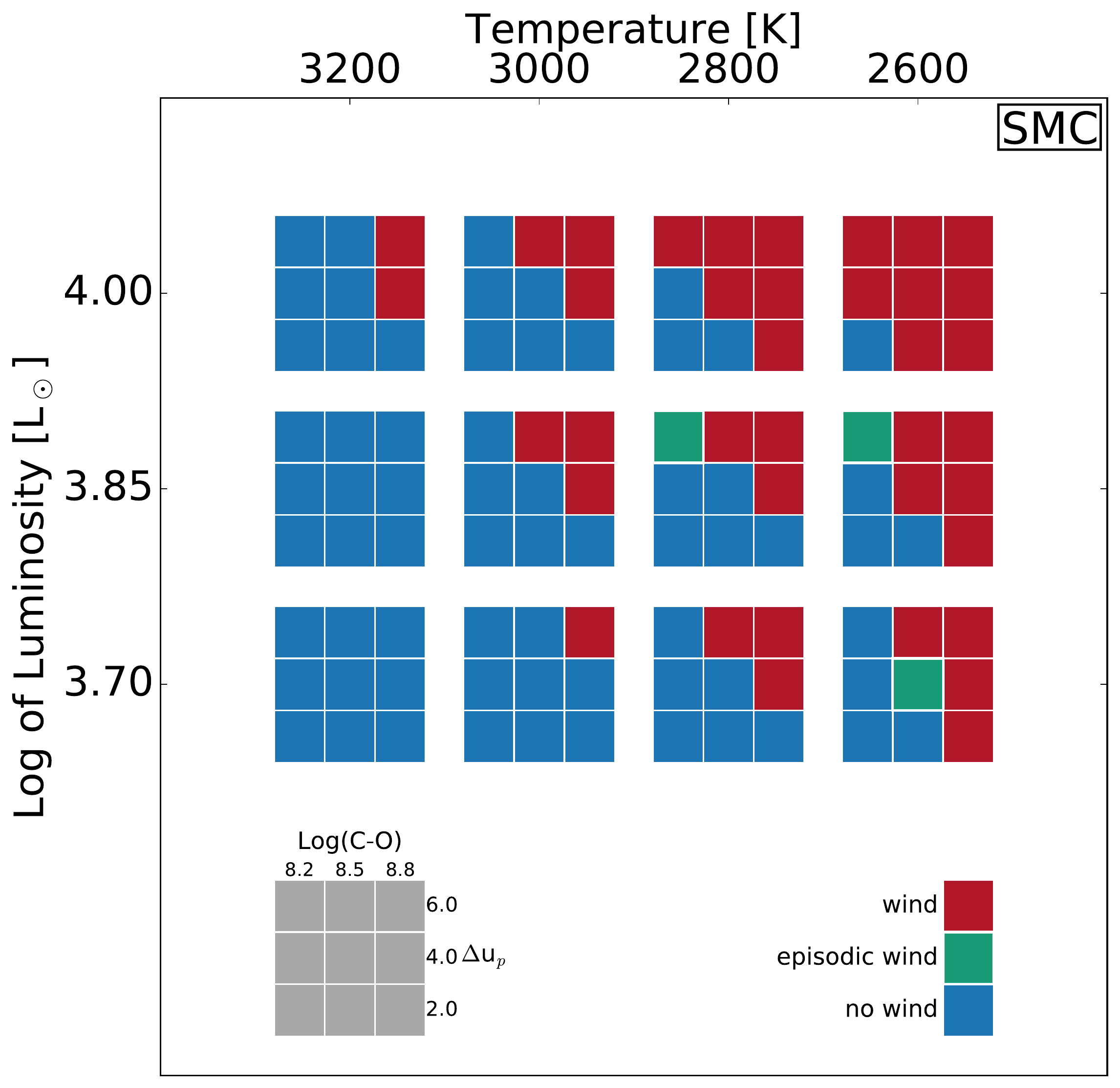}
\caption{Schematic overviews showing the dynamic behavior of solar-mass models for C-type AGB stars as a function of input parameters. The different panels, from top to bottom, show wind maps for models with solar metallicity ($[\mathrm{Fe/H}]=0$), LMC metallicity ($[\mathrm{Fe/H}]=-0.5$), and SMC metallicity ($[\mathrm{Fe/H}]=-1.0$). The red boxes represent models with a stellar wind, the blue boxes represent models with no wind, and the green boxes represent models with episodic mass loss. For each combination of luminosity and effective temperature the carbon excess and piston velocity are varied as indicated by the inset box.}
\label{f_windmaps}
\end{figure}

\section{Grid parameters}
\label{s_gridpar}
To explore the metallicity dependence of mass loss we calculate grids of DARWIN models at three different metallicities, corresponding to a solar-like environment and subsolar environments similar to the Large and Small Magellanic Clouds. For models at solar metallicity we assume solar abundances following \cite{asplund2005}, and for models at LMC and SMC metallicity we scale these solar abundances (except H, He, and C) by $-0.5$ dex and $-1.0$ dex, respectively. Consequently, all elements besides H, He, and C are scaled according to metallicity.

The raw material for the wind-driving dust species in carbon stars is the carbon that is not bound in CO molecules, i.e., carbon that instead can be found in molecular species such as C$_2$, C$_2$H, and C$_2$H$_2$. The most relevant quantity for dust formation, and, consequently, the mass loss in carbon stars is therefore the carbon excess. The carbon excess is defined as $\mathrm{C}-\mathrm{O}=\epsilon_C-\epsilon_O$, where $\epsilon_C$ and $\epsilon_O$ are the abundances of carbon and oxygen by number. Since carbon may be dredged up during the thermal pulses as AGB stars evolve, we keep the carbon abundance as a free parameter. This means that for every combination of stellar parameters, and for each metallicity, we compute models with carbon excesses of $\log(\mathrm{C-O})+12=8.2$, $8.5$, and $8.8$. Since the oxygen abundance is set by the overall metallicity, the C/O ratios for a given carbon excess will increase with decreasing metallicity. A summary of the metallicity; the abundances of carbon, oxygen, and nitrogen; the carbon excess; and the C/O ratios in the three different sets of DARWIN models can be seen in Table~\ref{t_opacity}. The relevance of the nitrogen abundance is discussed in Sect.~\ref{s_mol}.

The model grids are designed to cover typical stellar parameters (effective temperature, luminosity, and stellar mass) of observed carbon stars \citep[e.g.][]{ramstedt14}, as well as predicted values from stellar evolution models \citep[e.g.][]{marigo2017}. All models included have a current stellar mass of one solar mass. The grids cover effective temperatures of 2600\,K, 2800\,K, 3000\,K, and 3200\,K and stellar luminosities equal to $\log L_*/L_{\odot}=3.70$, 3.85, and 4.00. The pulsation period describing the variation at the inner boundary is calculated from the period-luminosity relation presented in \cite{feast1989}. The velocity amplitudes are set to $\Delta u_{\mathrm{p}}=2$, $4$, or $6$ km/s, resulting in shock amplitudes of about $15-20$ km/s in the inner atmosphere. The scaling factor $f_{\mathrm{L}}$ for the luminosity amplitude is set to 2, as proposed by \cite{eriksson14}. The combinations of input parameters covered by the three model grids (solar, LMC, and SMC) are listed in Table~\ref{t_grid}.

\begin{figure}
\centering
\includegraphics[width=0.9\linewidth]{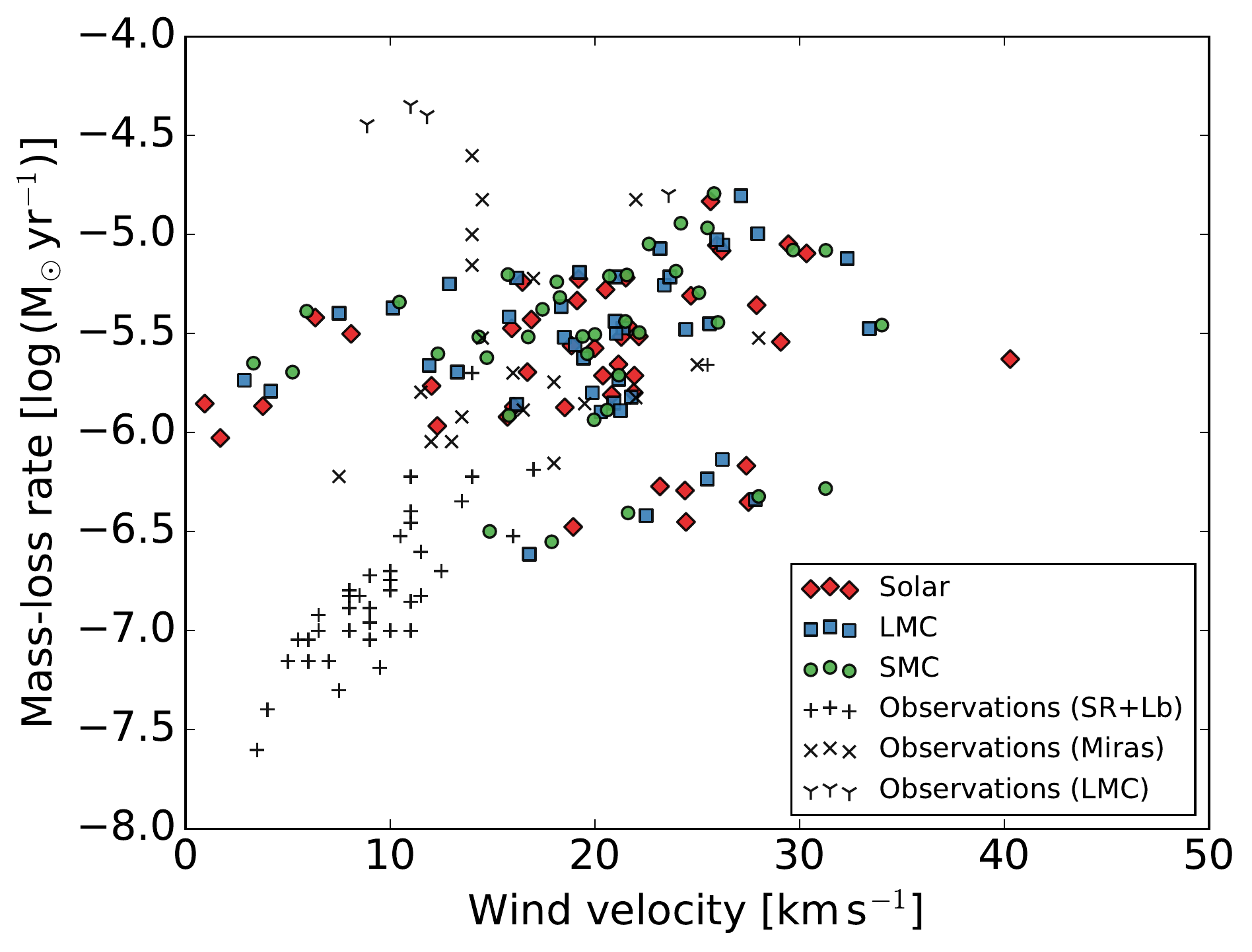}
\caption{Mass-loss rates vs. wind velocities for DARWIN models of carbon stars at different metallicities (solar, LMC, and SMC) and the corresponding observed properties for Galactic carbon stars and four sources from the LMC, derived from CO-lines \citep{schoier2001,ramstedt14, Groenewegen2016}.}
\label{f_dynall}
\end{figure}

\begin{figure}
\centering
\includegraphics[width=0.9\linewidth]{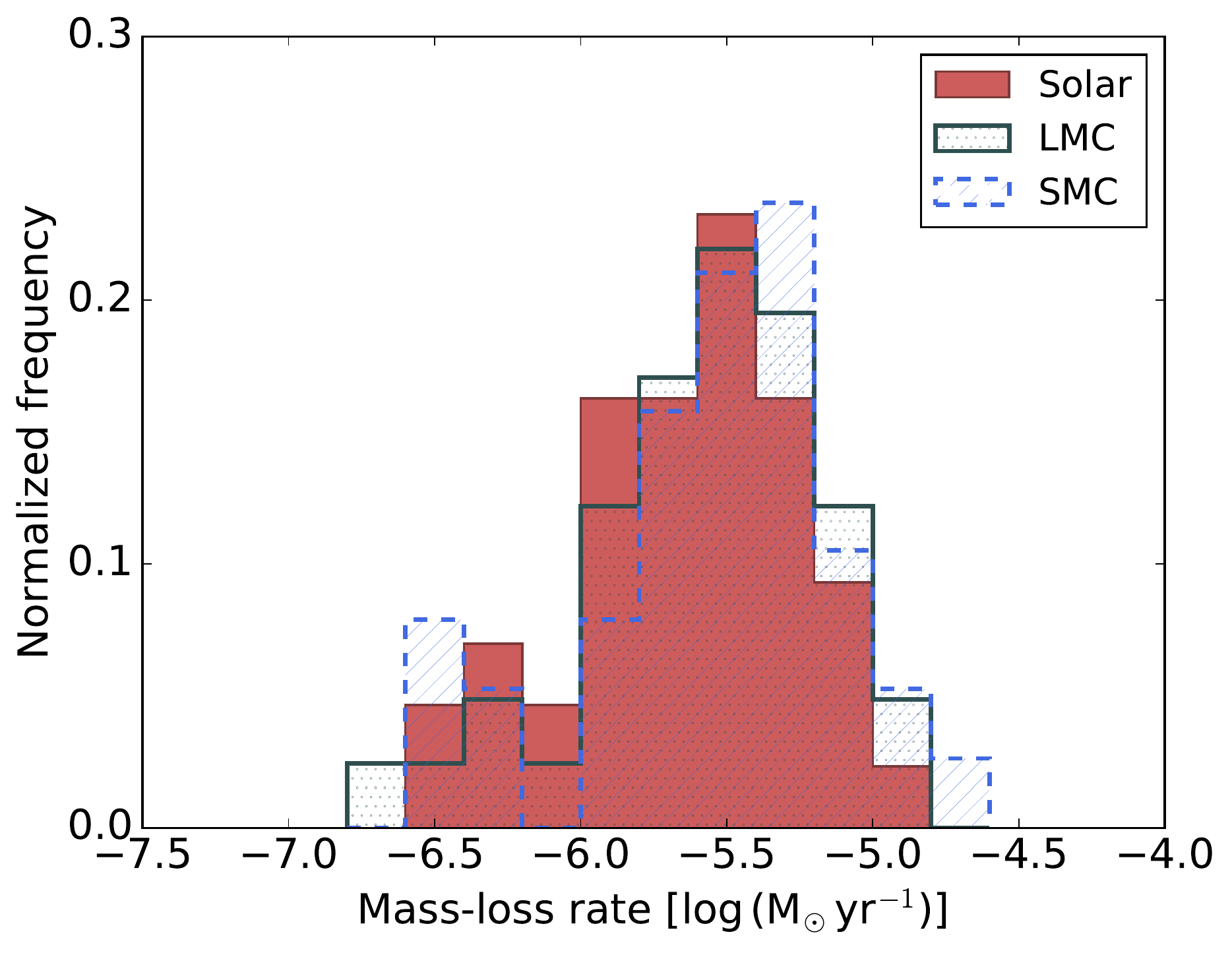}
\includegraphics[width=0.9\linewidth]{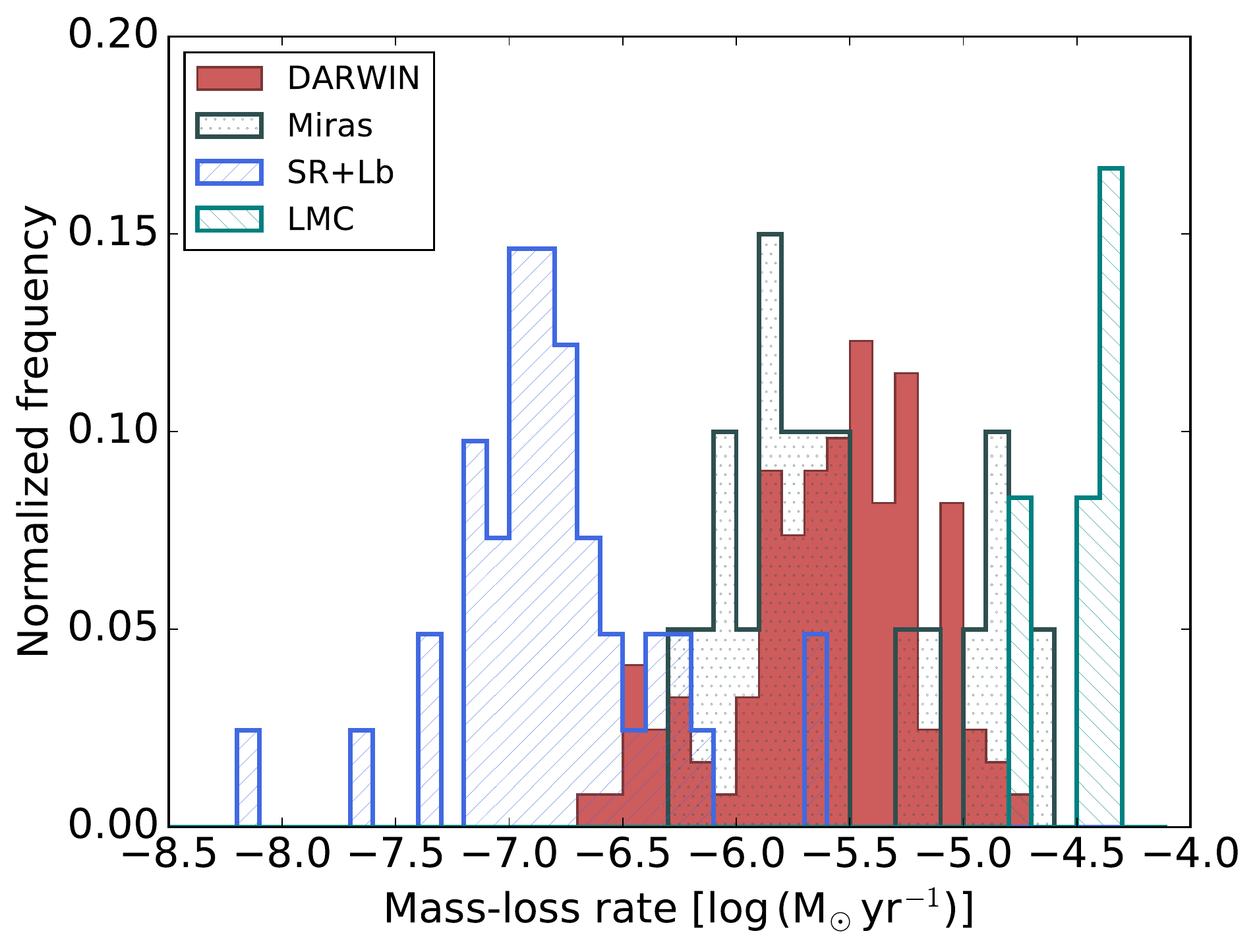}
\caption{Mass-loss rates for all models that develop a wind (indicated in red in Fig.~\ref{f_windmaps}). The upper panel shows model values sorted according to metallicity, whereas the lower panel shows both observational and model data. The frequency for the observational LMC data is divided by 3 to fit in the plot without losing details from the other observational sets.}
\label{f_histdyn}
\end{figure}

\begin{figure*}
\centering
\begin{tabular}{cc}
\includegraphics[width=0.45\textwidth]{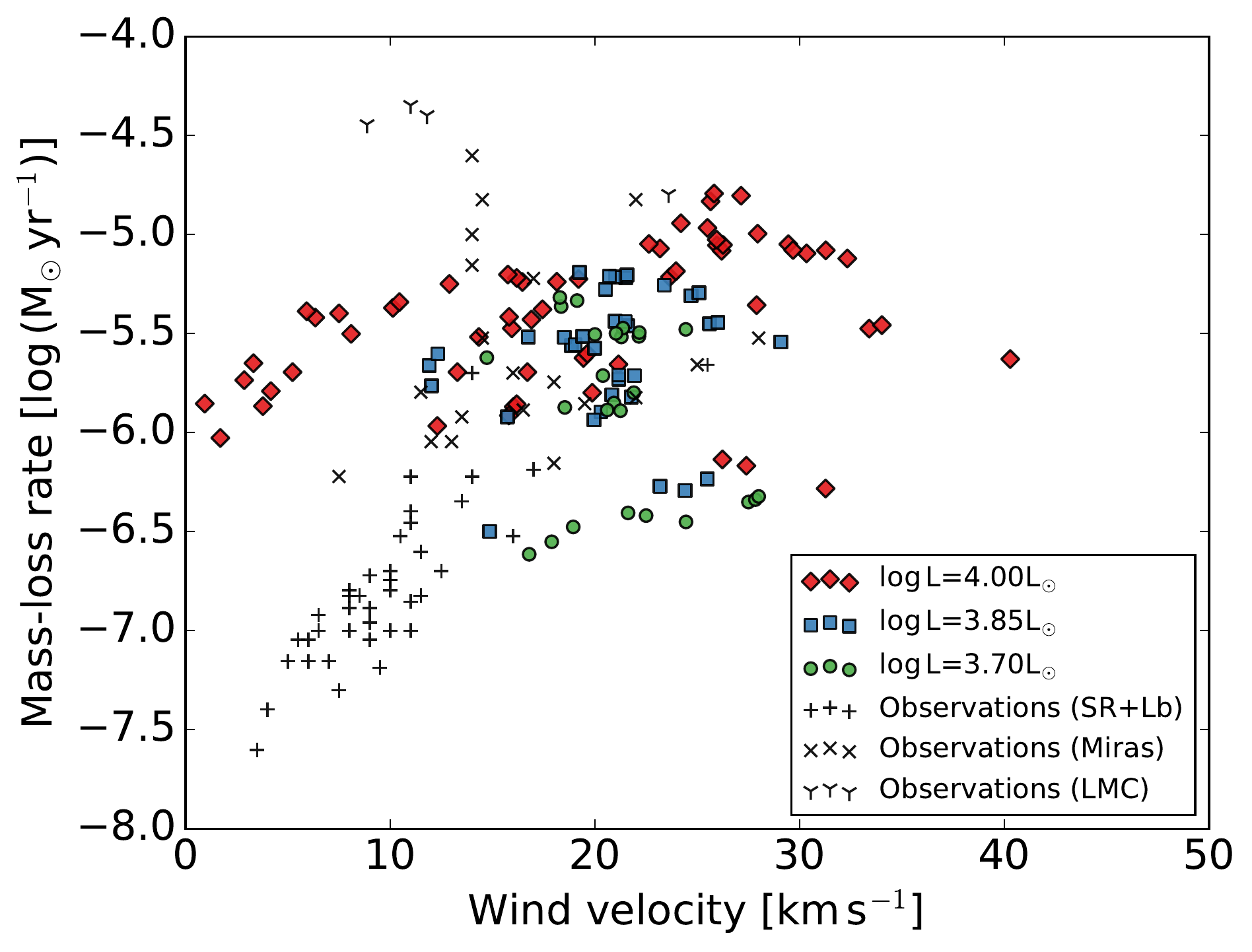} &
\includegraphics[width=0.45\textwidth]{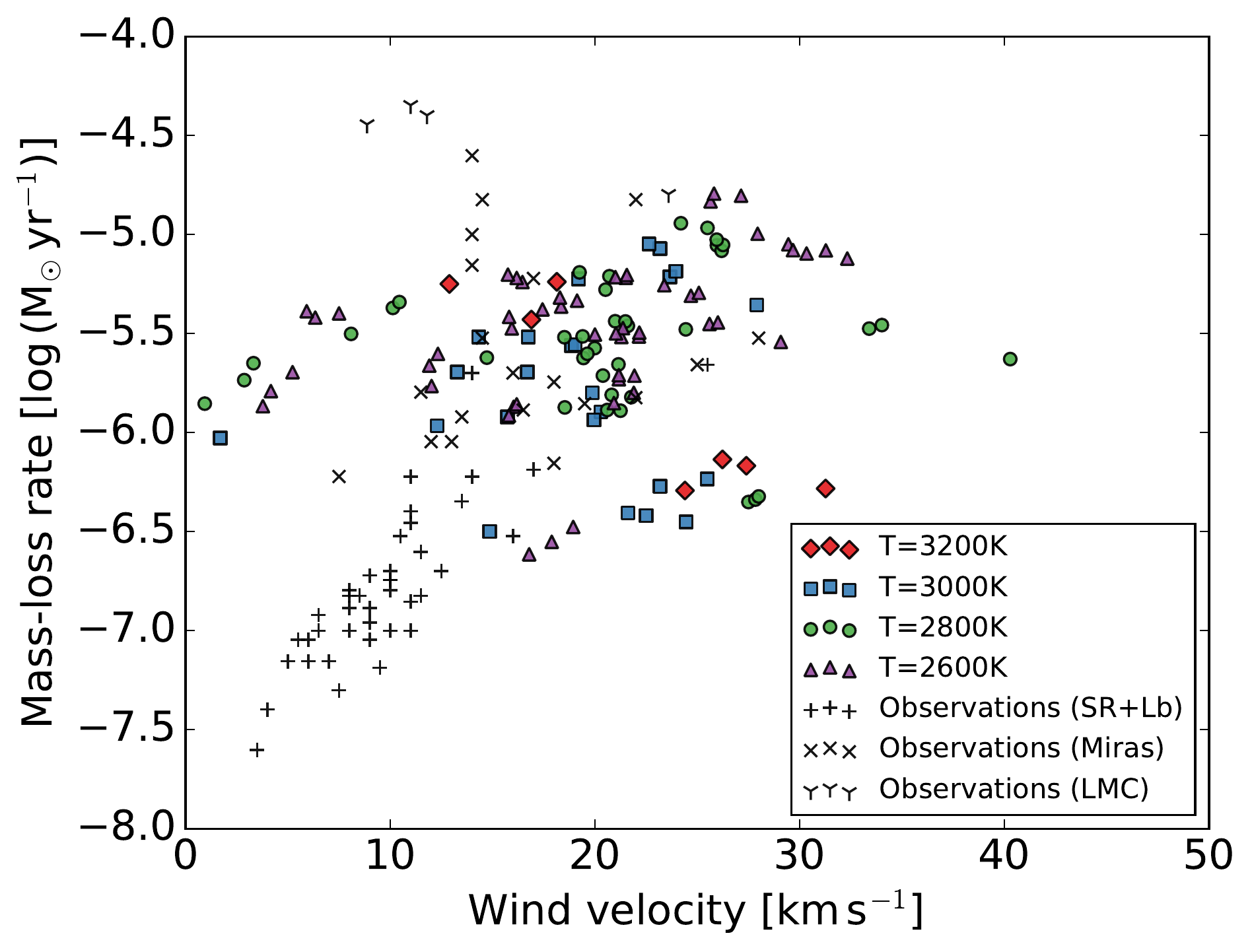}\\
\includegraphics[width=0.45\textwidth]{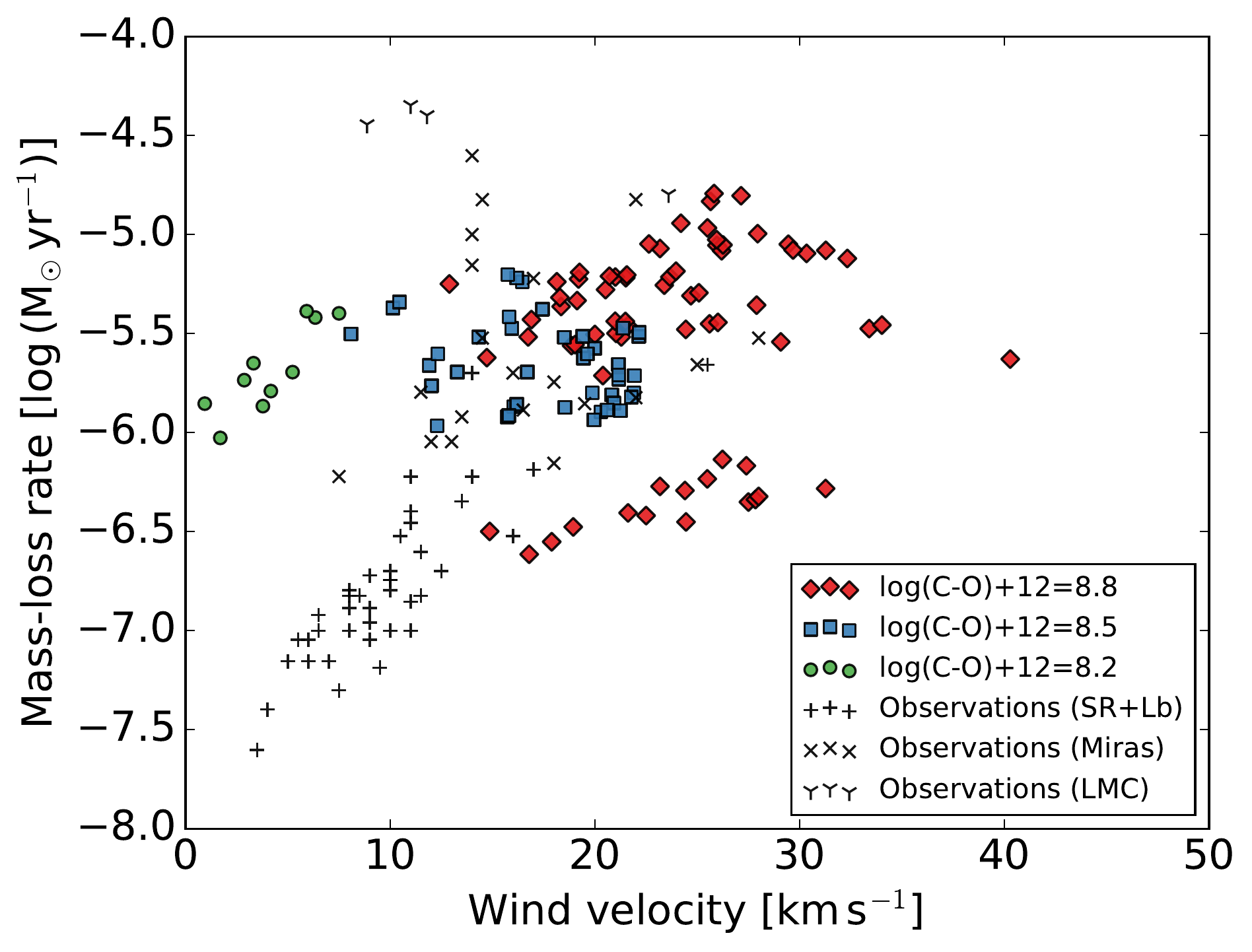}&
\includegraphics[width=0.45\textwidth]{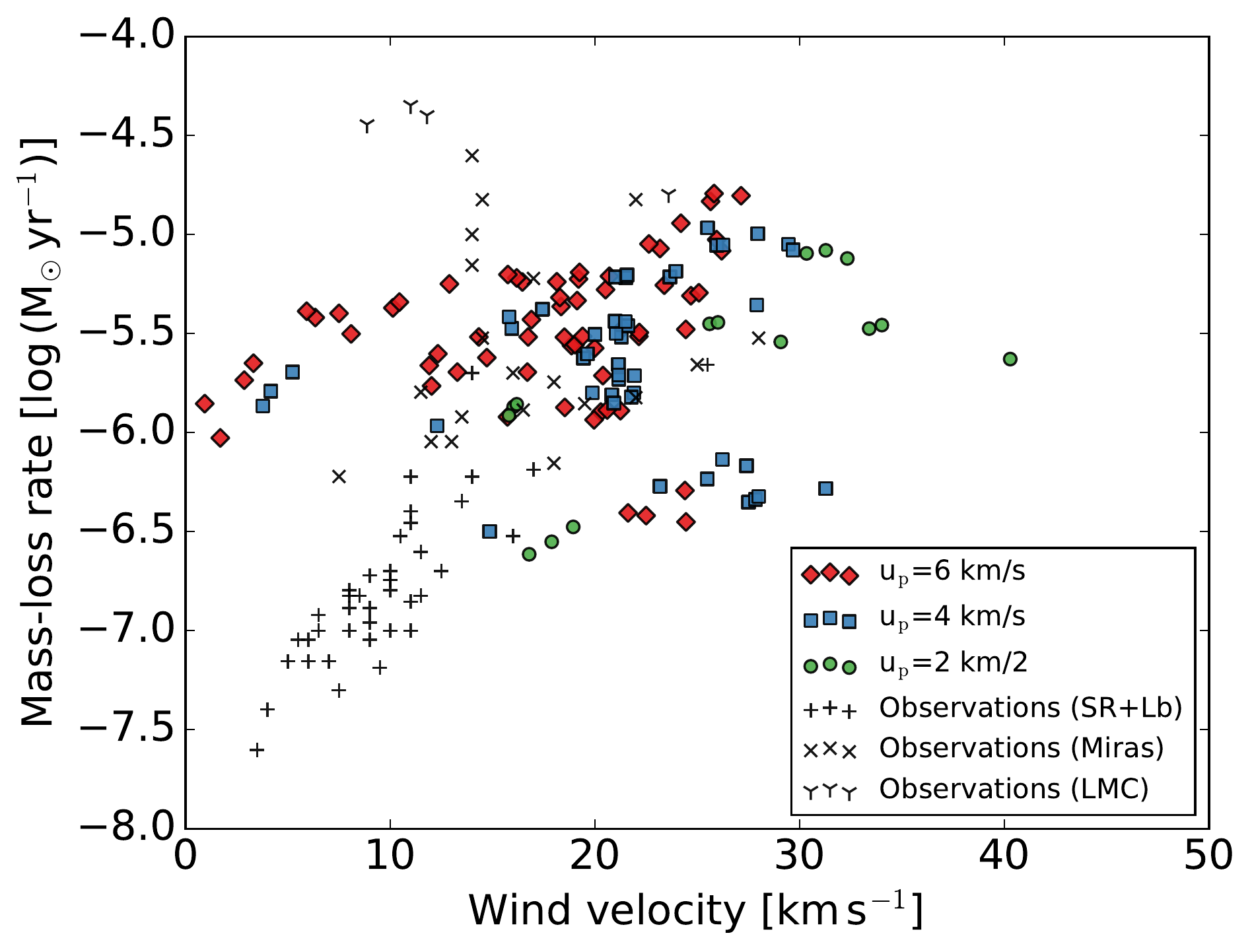} 
\end{tabular}
\caption{Mass-loss rates vs. wind velocities for DARWIN models of carbon stars at different metallicities (solar, LMC, and SMC) and the corresponding observed properties for Galactic carbon stars and four sources from the LMC, derived from CO-lines \citep{schoier2001,ramstedt14, Groenewegen2016}. All panels show the same data but with color-coding according to stellar luminosity (upper left), effective temperature (upper right), carbon excess (lower left), and piston velocity (lower right).}
\label{f_dynpar}
\end{figure*}

\section{Spectral synthesis and photometry}
\label{s_sed}
The DARWIN models consist of long time-series of snapshots of the atmosphere and wind structure. Each snapshot provides information about properties such as velocity, temperature, density, and grain size as a function of radial distance from the star. These atmospheric snapshots are further processed by a posteriori radiative transfer calculations to produce opacity sampling spectra ($R=10000$), using opacities from the COMA code \citep{aringer2016}. Based on these spectra we compute photometric filter magnitudes following the Bessell system \citep[described in][]{bess88,bess90}, and low-resolution spectra ($R=200$) covering the wavelength range between 0.33 and 25~$\mu$m. Mean visual and near-IR magnitudes, as well as mean molecular concentrations, are calculated from 30-60 atmospheric snapshots, selected to be equidistant in phase and covering three consecutive pulsation cycles (see Fig.~\ref{f_timeseries}). 

The treatment of gas and dust opacities is consistent in DARWIN and COMA, and includes the molecular species CO, CH, C$_2$, SiO, CN, TiO, H$_2$O, C$_2$H$_2$, HCN, C$_3$, OH, VO, CO$_2$, SO$_2$, HF, HCl, CH$_4$, FeH, CrH, ZrO, YO, CS, MgH, CaH, and TiH. The optical data for amorphous carbon are taken from \cite{rolmar91}.

\section{Dynamical results}
\label{s_dynres}
The wind properties of the DARWIN models are calculated by extracting the mass-loss rates and wind velocities at the outermost layers and averaging over typically a few hundred pulsation periods. The early pulsation periods are excluded to avoid transient effects from ramping up the amplitude. The top panel of Fig.~\ref{f_timeseries} shows an example of such a time-series for a model with input parameters $M_*=1\,\mathrm{M}_{\odot}$, $\log L_*/L_{\odot}=4.00$, $T_*=2600\,$K, $u_{\mathrm{p}}=6$\,km/s, $\log(\mathrm{C-O})+12=8.5$, and $[\mathrm{Fe/H}]=-1.0$ (SMC). The sections in black mark the time interval over which the mass-loss rate and wind velocity are averaged, and the two embedded panels show the snapshots selected for the a posteriori radiative transfer (see Sect.~\ref{s_mol}).

\subsection{Schematic overview of the dynamical properties}
\label{s_windmaps}
An overview of the dynamical properties of DARWIN models at different metallicities is shown in Fig.~\ref{f_windmaps}. The different panels, from top to bottom, illustrate wind properties for models (with a current mass of one solar mass) with solar, LMC, and SMC metallicity, respectively. These "windmaps" are organized like an HR-diagram, with decreasing temperature in the positive x-direction and increasing luminosity in the positive y-direction, and each individual box depicted represents a model. The boxes are arranged in subsets, where each subset corresponds to a combination of effective temperature and stellar luminosity. The boxes within each subset are organized so that the parameter describing the amplitude of the velocity variation at the inner boundary (the piston velocity $\Delta u_{\mathrm{p}}$) increases upward and the parameter describing the amount of free carbon available for grain formation (the carbon excess $\log(\mathrm{C-O})+12$) increases toward the right. 

The dynamical properties of each model, or box, are indicated by the color-coding. Red boxes represent models that develop a stellar wind, blue boxes represent models with no wind, and green boxes indicate models with episodic mass loss. Consequently, these wind maps illustrate the combinations of stellar parameters (mass, luminosity, and effective temperature) that produce outflows. They also indicate where the boundary between models with and without a wind is situated, in terms of stellar parameters. By studying these wind maps it is clear that mass loss is facilitated by high luminosities, low effective temperatures, and high carbon excess, at both solar and subsolar metallicities. Another thing to note is that the wind/no-wind boundary slightly shifts as the metallicity decreases. The subset of boxes corresponding to the stellar parameters $\log L_*/L_{\odot}=4.00$ and $T_*=3000\,$K is an example of this. At solar metallicities, DARWIN models with a carbon excess as low as $\log(\mathrm{C-O})+12=8.2$ can produce stellar winds, whereas models at lower metallicities require higher carbon excess and higher piston velocities to produce outflows. Similar shifts are seen in the subsets corresponding to the stellar parameters $\log L_*/L_{\odot}=3.85$ and $T_*=2800\,$K and $\log L_*/L_{\odot}=3.70$ and $T_*=2600\,$K, but then only between LMC and SMC metallicity.

To summarize, the parameter space covered by models with and without wind is quite similar at different metallicities. However, it becomes slightly smaller, when the metal content decreases (see Sect.~\ref{s_mol}).

\subsection{Trends in mass-loss rate with metallicity}
The mass-loss rates and wind velocities for all models that produce a stellar wind are shown in Fig.~\ref{f_dynall}, color-coded according to metallicity. As a comparison we also plot observed wind properties derived from observations of CO-line emission for a volume-limited sample of nearby Galactic carbon stars \citep{schoier2001,ramstedt14} and four carbon stars in the LMC \citep{Groenewegen2016}. As can be seen in Fig.~\ref{f_dynall}, there is no systematic trend in the wind properties calculated from the DARWIN models with respect to metallicity. This is further confirmed by the histogram in the top panel of Fig.~\ref{f_histdyn}, showing the mass-loss rates from the DARWIN models, sorted by metallicity. 

We divide the observational dataset for the nearby carbons stars into two subsets, consisting of Miras and Semiregulars/Irregulars. The lower panel of Fig.~\ref{f_histdyn} shows a histogram over the mass-loss rates from all DARWIN models, together with the observed mass-loss rates from Galactic Miras and Semiregulars/Irregulars and the small sample of carbon stars in LMC. It should be kept in mind that most of the observed values are from nearby Galactic carbon stars and that the model values are derived from wind models at different metallicities. Furthermore, in the model grid every point has equal weight, which does not reflect how probable each combination of stellar parameters is in a population of real carbon stars. Therefore, the frequency of the observational data and models values in the histograms should not be compared directly. Even so, it is clear that the models do not reproduce the observed data from Semiregulars and Irregulars, but cover most of the observed range from the Galactic Miras, except for carbon stars with very high mass-loss rates. Such high mass-loss rates can probably be reached by including models with higher luminosity or lower current mass (less gravitational potential for the radiative acceleration to overcome). The variability at the inner boundary of these models are designed to mimic Miras (see Sect.~\ref{s_puls}), which could explain why they do not reproduce the low mass-loss rates of Semiregulars and Irregulars. 

The dynamical properties of all DARWIN models are listed in Table~\ref{t_dynprop}. The first  four columns list selected input parameters of each model (luminosity, effective temperature, carbon excess, and piston velocity). The next six columns list the dynamical properties (mass-loss rate and wind velocity) at the three different metallicities (solar, LMC, and SMC). The last three columns show the relative difference in mass-loss rate between models with different metallicities, but otherwise the same input parameters, e.g., $\log\dot{M}_{\mathrm{solar}}-\log\dot{M}_{\mathrm{LMC}}$.

\begin{figure}
\centering
\includegraphics[width=0.9\linewidth]{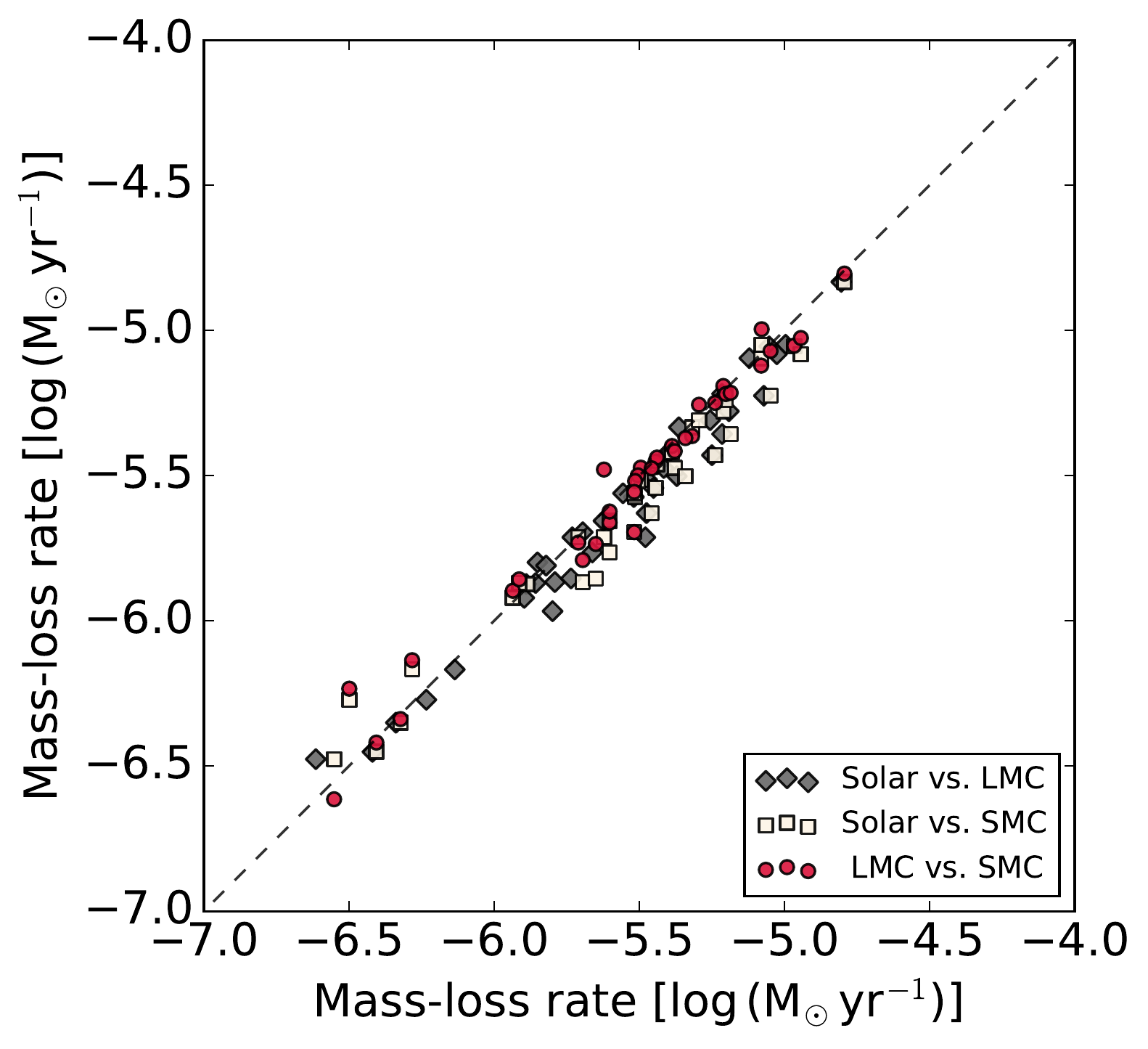}
\includegraphics[width=0.9\linewidth]{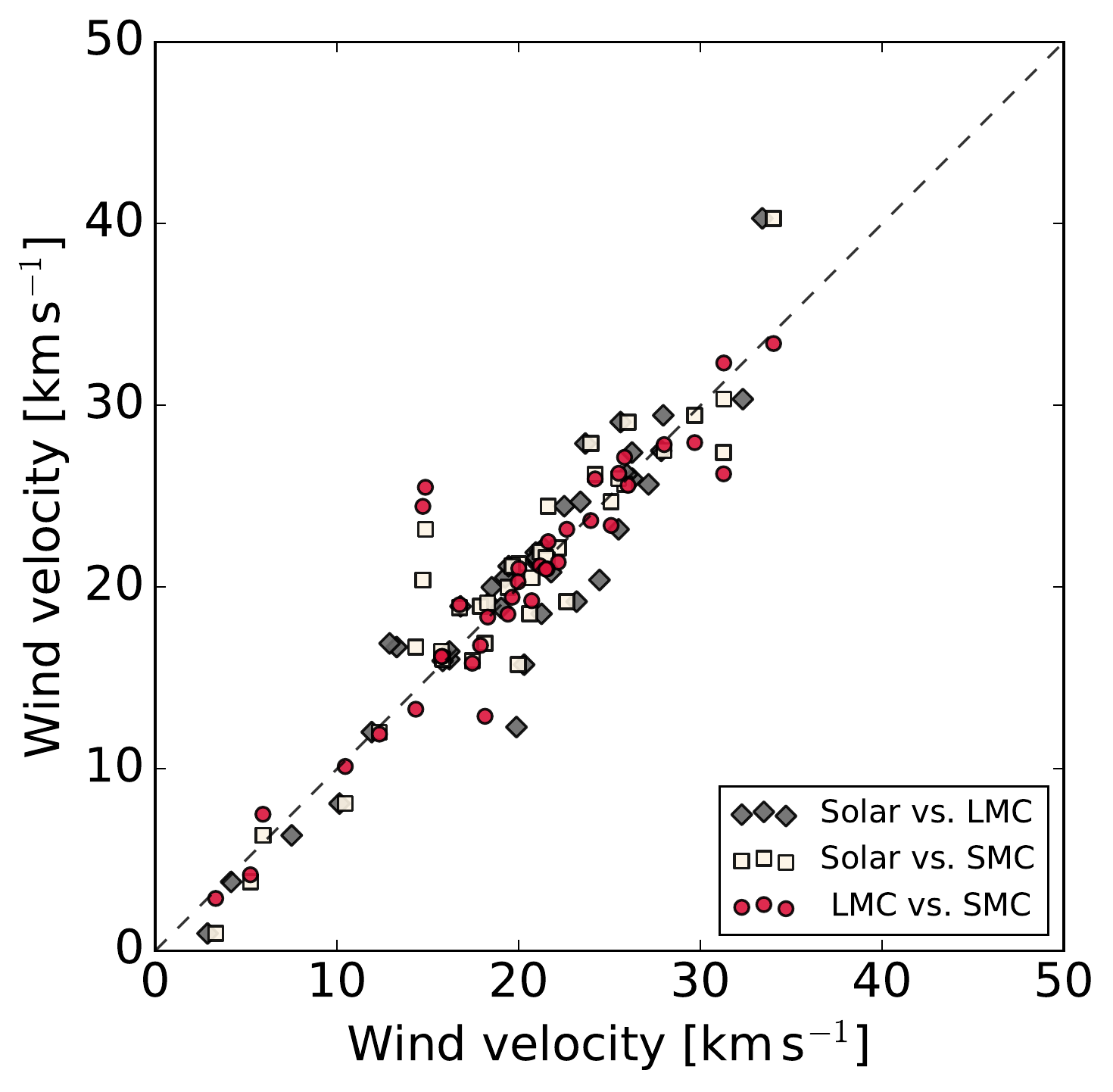}
\caption{Mass-loss rates (upper panel) and wind velocities (lower panel) from models with the same input parameters, but different metallicities, plotted against each other. The gray diamonds show wind properties from models with solar metallicity vs. LMC metallicity, the white squares show wind properties from models with solar metallicity vs. SMC metallicity, and the red circles show wind properties from models with LMC metallicity vs. SMC metallicity.}
\label{f_dyncomp}
\end{figure}

\subsection{Trends in mass loss vs. stellar parameters}
\label{s_trendpar}
Figure~\ref{f_dynpar} shows the wind properties color-coded according to luminosity, effective temperature, carbon excess, and piston velocity. The trends in mass-loss rates and wind velocities with respect to these input parameters are similar to what was found in previous studies for carbon stars at solar metallicity \citep[e.g.][]{mattsson2010,eriksson14}: a higher luminosity correlates with an increased mass-loss rate and a higher carbon excess results in higher wind velocity. 


It is important to remember that not all combinations of stellar parameters are equally probable. Figure~\ref{f_dynall} shows a cluster of model values in the top left corner ($v<10$\,km/s and $\dot{M}>10^{-6}$\,M$_{\odot}$/yr) and in the lower middle ($v>15$\,km/s and  $\dot{M}<10^{-6.5}$\,M$_{\odot}$/yr) where stars are not observed. From Fig.~\ref{f_dynpar} we can deduce that models that combine a high luminosity ($\log L_*/L_{\odot}=4.00$) with a low carbon excess ($\log(\mathrm{C-O})+12=8.2$) do not produce wind properties compatible with observed values from nearby Galactic carbon stars. Models in the middle cluster all have a high carbon excess, but otherwise different combinations of luminosities, effective temperatures, and piston velocities that are close in parameter space to the wind/no-wind boundary in Fig.~\ref{f_windmaps}. As carbon stars evolve on the AGB, the effective temperature and current mass decrease and the carbon excess increases, pushing them more firmly toward stellar parameters that facilitate high mass-loss rates. The fact that such stars are not observed in the Galactic sample of carbon stars suggests that having stellar parameters on the wind/no-wind boundary and a high carbon excess is a very short-lived evolutionary phase.

The wind properties for three of the carbon stars in the LMC sample stand out by exhibiting a combination of high mass-loss rates and low wind velocities (the high mass-loss rates observed in these stars are further confirmed by estimates based on dust emission, see \cite{loon1999} and \cite{Groenewegen2007}). The estimated luminosities of these three stars are $\log L_*/L_{\odot}=3.75$, 3.98 and 4.19 \citep{Groenewegen2016}, i.e., they are not all extremely luminous. However, if these stars have a low current mass this could lead to high mass-loss rates even at more moderate luminosities. As mentioned above, the combination of high mass-loss rates and low wind velocities is not seen in the Galactic observational sample. This could be a selection effect, as the four stars are very dust enshrouded, with pulsation periods of approximately 1100 days, whereas the stars in the Galactic sample all have periods shorter than 700 days (the DARWIN models included in this study all have periods shorter than 550 days). According to evolutionary models of the TP-AGB phase by \cite{marigo2013}, these four carbon stars in the LMC are probably experiencing their last thermal pulses, an evolutionary phase when intense mass loss is stripping away the stellar envelope \citep[see Fig.~3 in][]{Groenewegen2016}. Such extreme AGB stars might simply not be present in the Galactic sample of nearby field carbon stars.

\subsection{Comparison of wind properties at different metallicities}
\label{s_compdyn}
A comparison between mass-loss rates and wind velocities produced by DARWIN models with the same input parameters but different chemical composition is shown in Fig.~\ref{f_dyncomp}. In this figure the wind properties at solar metallicity are plotted against the wind properties at LMC and SMC metallicities (gray diamonds and white squares), and similarly the wind properties at LMC metallicity are plotted against the wind properties at SMC metallicity (red circles). If models with different metallicities, but otherwise the same input parameters, produce similar wind properties, they will end up close to the dashed line. As can be seen, both the mass-loss rates and wind velocities line up close to the dashed diagonal (the mass-loss rates are plotted in logarithmic scale whereas the wind velocities are plotted in linear scale). The relative difference in mass-loss rates between models with different metallicities but the same input parameters is on average approximately 15\%, and at most around 60\% (see Cols.~11-13 in Table~\ref{t_dynprop}). 

Figure~\ref{f_dyncomp} also shows that SMC and LMC models in general have slightly higher mass-loss rates compared to solar models. This trend is also seen in the top panel of Fig.~\ref{f_histdyn}, where the SMC and LMC models have a slightly higher frequency at the highest mass-loss rates. The slightly higher mass-loss rates in the subsolar models suggest that the density in the wind acceleration zone is slightly increased at lower metallicities, which in turn could be a consequence of small differences in the atmospheric structure due to metallicity. Nevertheless, these differences in mass-loss rates are very small compared to the uncertainty of observed mass-loss rates, which are estimated to be as high as a factor of three \citep{Ramstedt2008}.

\subsection{Grain properties}

\begin{figure}
\centering
\includegraphics[width=0.9\linewidth]{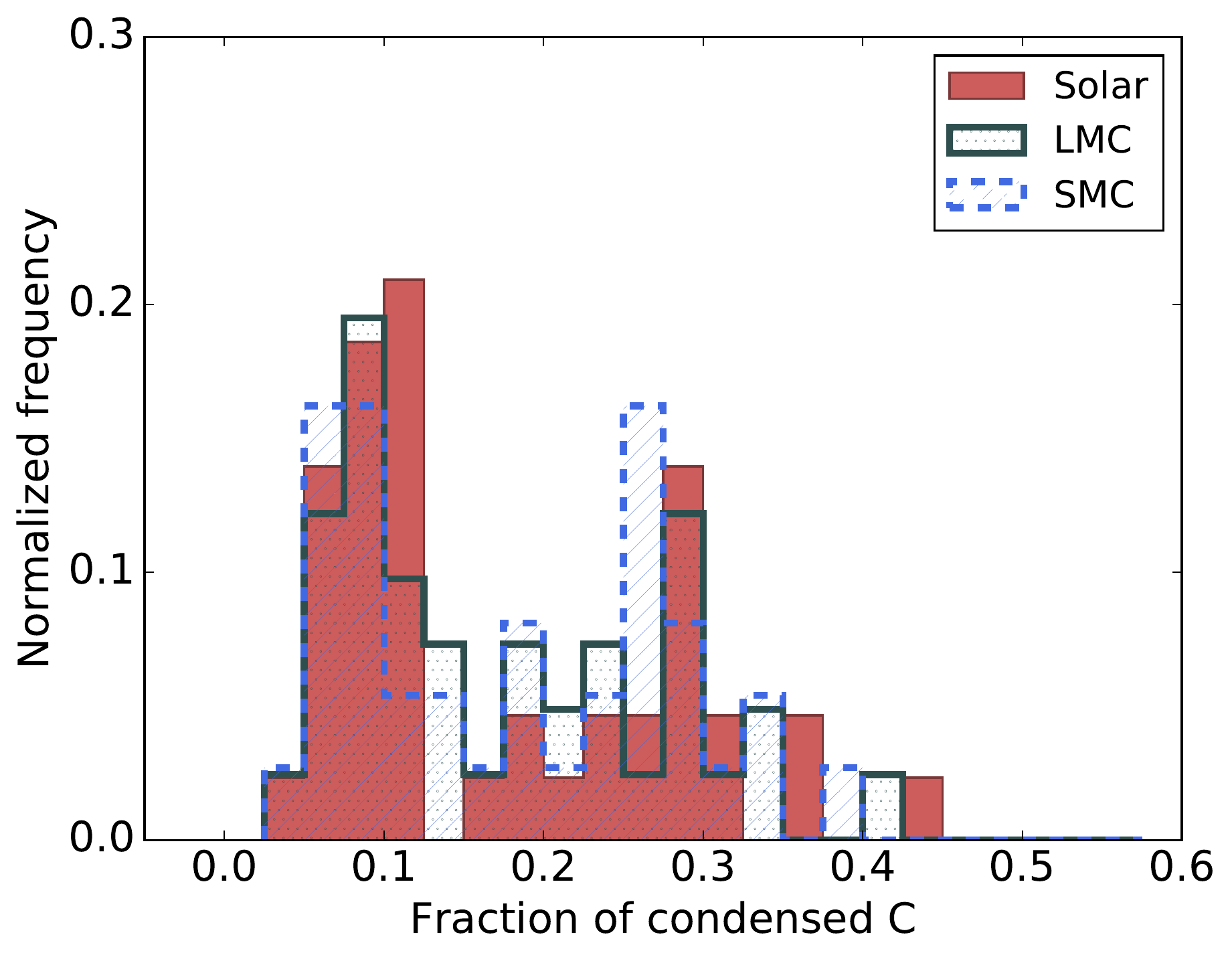} 
\includegraphics[width=0.9\linewidth]{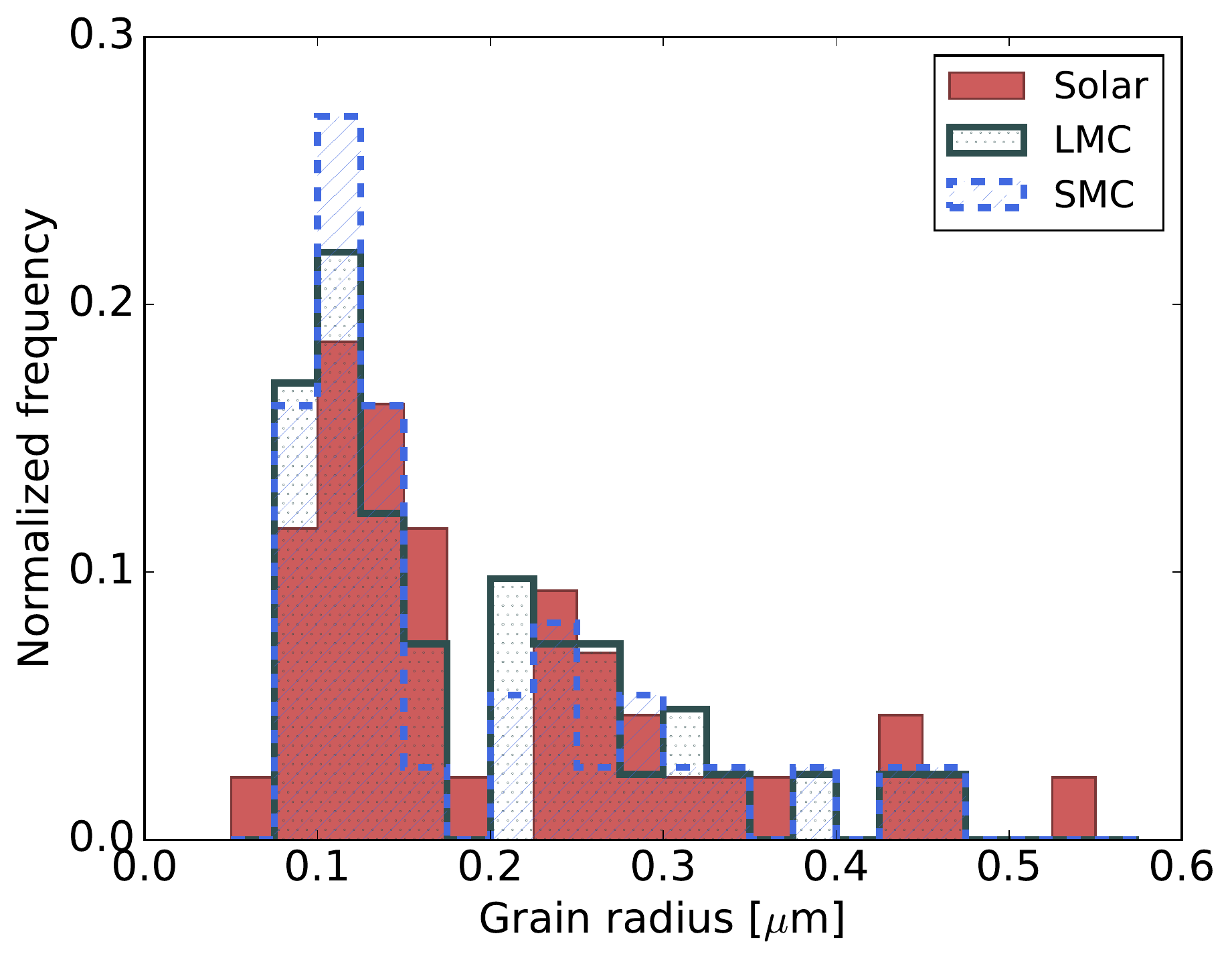}
\includegraphics[width=0.9\linewidth]{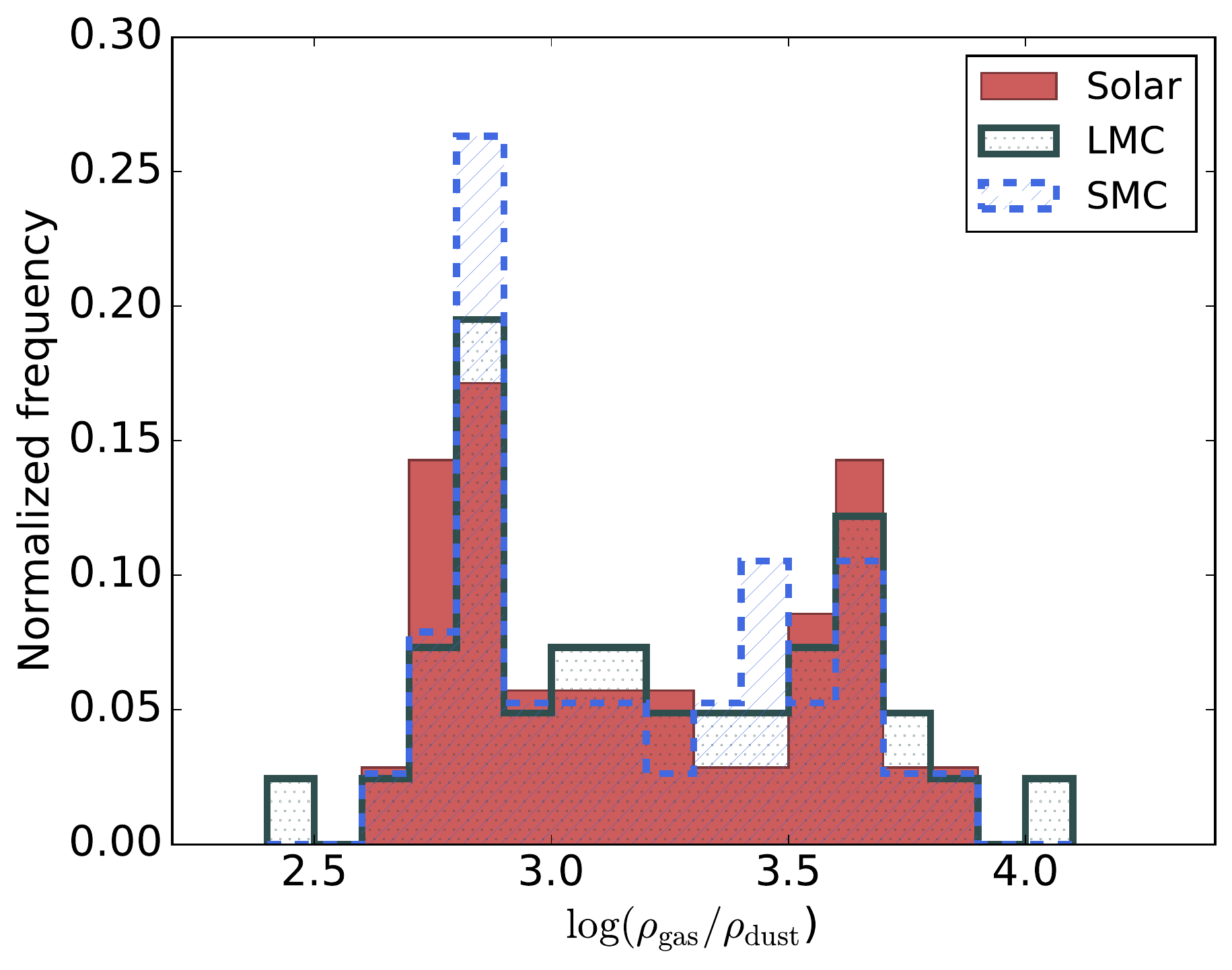}
\caption{Degree of condensed carbon (upper panel), grain sizes (middle panel) and gas-to-dust mass ratios (lower panel) in the outermost atmospheric layers, averaged over time, for all models that develop a wind (indicated in red in Fig.~\ref{f_windmaps}). The histograms are color-coded according to different metallicities.}
\label{f_grain}
\end{figure}

Figure~\ref{f_grain} shows histograms of the degree of condensed carbon, the average grain sizes, and gas-to-dust mass ratios for the DARWIN models that produce outflows, sorted according to metallicity. Similar to the wind properties, there are no large systematic difference with respect to metallicity concerning the distribution of grain sizes, how much carbon is condensed, or the gas-to dust ratios. For both solar and subsolar metallicities the typical values for the grain sizes are $0.1-0.5\,\mu$m, the degree of condensed carbon varies between 5\% and 40\% and the gas-to-dust ratios range between 500 and 10000. These results agree reasonably well with the observational estimates by \cite{Ramstedt2008} and \cite{nanni2018}, who found dust-to-gas mass ratios between 500 and 3000. This is, however, significantly higher than the standard value of 200 used in many studies that derive mass-loss rates from SED fitting \citep[e.g.][]{boyer2012,gullieuszik2012,srinivasan2016}.

As mentioned in previous sections, the grain sizes presented here are the result of a parameter study of dynamical wind models, not a stellar population study. Population studies constraining the dust properties for Galactic sources and the Magellanic Clouds \citep[e.g.][]{nanni2016,agli2015a,agli2015b,agli2017} seem to favor smaller carbon grains. For example, \cite{nanni2016} find that the near-infrared and mid-infrared colors in the SMC are best reproduced by amorphous carbon grains of sizes $0.035-0.12\,\mu$m, whereas \cite{agli2015a} find carbon grains of sizes $0.05-0.2\,\mu$m when constraining the dust properties in the LMC. These values are near the lower end of the range produced by our models.


It should be noted, however, that these stellar populations studies are based on stationary wind models that ignore the time-dependent effects of pulsation and shocks, i.e., these models only describe the dust-driven part of the pulsation-enhanced dust-driven outflows. A full treatment of the wind mechanism requires that we consider the time-dependence of both gas dynamics and dust formation as there is a significant feedback between these two processes. The density enhanced layers created by the pulsation-induced shock waves play a significant role for the efficiency of grain growth; in turn, the radiation pressure on the newly formed dust particles strongly influences the dynamics, and consequently the atmospheric structure. Furthermore, stationary wind models require a priori assumptions about the mass-loss rates, whereas DARWIN models predict both mass-loss rates and grain sizes self-consistently.

\begin{table*}[!htbp]
\caption{Dynamical properties of DARWIN models for C-type AGB stars of solar mass at three different metallicities (solar, LMC, and SMC).}             
\label{t_dynprop}      
\centering          
\begin{tabular}{c c c c | r r | r r | r r | r r r}     
\hline\hline       
\multicolumn{4}{c}{Model parameters} &  \multicolumn{2}{c}{Solar} & \multicolumn{2}{c}{LMC}	& \multicolumn{2}{c}{SMC} & \multicolumn{3}{c}{$\Delta \log\dot{M_{ij}}$}\\
\hline  
$\log L_*/L_{\odot}$ & $T_*$ & [C-O] & $u_\mathrm{p}$ & $\langle\dot{M_1}\rangle$ & $\langle u_1\rangle$ & $\langle\dot{M_2}\rangle$ & $\langle u_2\rangle$ & $\langle\dot{M_3}\rangle $ & $\langle u_3\rangle$ & $\dot{M}_{12}$ & $\dot{M}_{13}$ & $\dot{M}_{23}$\\ 
        & $\mathrm{[K]}$ & & $\mathrm{[km/s]}$ & $\mathrm{[M_{\odot}/yr]}$ & $\mathrm{[km/s]}$ & $\mathrm{[M_{\odot}/yr]}$ & $\mathrm{[km/s]}$ & $\mathrm{[M_{\odot}/yr]}$ & $\mathrm{[km/s]}$\\  
\hline                    
3.70 & 2600 & 8.5 & 4 & 1.6e-06 & 21.9 & 1.4e-06 & 20.9 &         	&        &  $0.06$  &              &     \\
3.70 & 2600 & 8.5 & 6 & 3.1e-06 & 22.2 & 3.4e-06 & 21.4 & 3.2e-06 & 22.2 & $-0.04$ & $-0.01$ & $ 0.03$\\
3.70 & 2600 & 8.8 & 2 & 3.3e-07 & 18.9 & 2.4e-07 & 16.8 & 2.8e-07 & 17.9 & $ 0.14$ & $ 0.07$ & $-0.07$\\
3.70 & 2600 & 8.8 & 4 & 3.0e-06 & 21.3 & 3.2e-06 & 21.0 & 3.1e-06 & 20.0 & $-0.03$ & $-0.01$ & $ 0.01$\\
3.70 & 2600 & 8.8 & 6 & 4.6e-06 & 19.1 & 4.3e-06 & 18.4 & 4.8e-06 & 18.3 & $ 0.03$ & $-0.02$ & $-0.05$\\
3.70 & 2800 & 8.5 & 6 & 1.3e-06 & 18.5 & 1.3e-06 & 21.3 & 1.3e-06 & 20.6 & $ 0.00$ & $ 0.00$ & $ 0.00$\\
3.70 & 2800 & 8.8 & 4 & 4.5e-07 & 27.5 & 4.6e-07 & 27.8 & 4.9e-07 & 25.0 & $-0.01$ & $-0.04$ & $-0.03$\\
3.70 & 2800 & 8.8 & 6 & 1.9e-06 & 20.4 & 3.3e-06 & 24.4 & 2.8e-06 & 21.3 & $-0.24$ & $-0.17$ & $ 0.07$\\
3.70 & 3000 & 8.8 & 6 & 3.5e-07 & 24.5 & 3.8e-07 & 22.5 & 3.9e-07 & 21.6 & $-0.04$ & $-0.05$ & $-0.01$\\
\hline
3.85 & 2600 & 8.5 & 4 & 1.9e-06 & 21.9 & 1.9e-06 & 21.2 & 1.9e-06 & 21.2 & $ 0.00$ & $ 0.00$ & $ 0.00$\\
3.85 & 2600 & 8.5 & 6 & 1.7e-06 & 12.0 & 2.2e-06 & 11.9 & 2.5e-06 & 12.3 & $-0.11$ & $-0.17$ & $-0.06$\\
3.85 & 2600 & 8.8 & 2 & 2.9e-06 & 29.1 & 3.5e-06 & 25.6 & 3.6e-06 & 26.0 & $-0.08$ & $-0.09$ & $-0.01$\\
3.85 & 2600 & 8.8 & 4 & 6.1e-06 & 21.5 & 6.1e-06 & 21.0 & 6.3e-06 & 21.6 & $ 0.00$ & $-0.01$ & $-0.01$\\
3.85 & 2600 & 8.8 & 6 & 4.9e-06 & 24.7 & 5.6e-06 & 23.4 & 5.1e-06 & 25.1 & $-0.06$ & $-0.02$ & $ 0.04$\\
3.85 & 2800 & 8.5 & 4 & 1.6e-06 & 20.8 & 1.5e-06 & 21.8 &              &         &  $0.03$ &               &   \\
3.85 & 2800 & 8.5 & 6 & 2.7e-06 & 20.0 & 3.0e-06 & 18.5 & 3.1e-06 & 19.4 & $-0.05$ & $-0.06$ & $-0.01$\\
3.85 & 2800 & 8.8 & 4 & 3.5e-06 & 21.6 & 3.7e-06 & 21.0 & 3.6e-06 & 21.5 & $-0.02$ & $-0.01$ & $ 0.01$\\
3.85 & 2800 & 8.8 & 6 & 5.3e-06 & 20.5 & 6.4e-06 & 19.3 & 6.2e-06 & 20.7 & $-0.08$ & $-0.07$ & $ 0.01$\\
3.85 & 3000 & 8.5 & 6 & 1.2e-06 & 15.7 & 1.3e-06 & 20.3 & 1.2e-06 & 20.0 & $-0.03$ & $ 0.00$ & $ 0.03$\\
3.85 & 3000 & 8.8 & 4 & 5.4e-07 & 23.2 & 5.8e-07 & 25.5 & 3.2e-07 & 14.9 & $-0.03$ & $ 0.22$ & $ 0.26$\\
3.85 & 3000 & 8.8 & 6 & 2.8e-06 & 18.9 & 2.8e-06 & 19.0 & 3.0e-06 & 16.7 &  $0.00$ & $-0.03$ & $-0.03$\\
3.85 & 3200 & 8.8 & 6 & 5.1e-07 & 24.4 &              &         &              &         &              &               &     \\ 
\hline
4.00 & 2600 & 8.2 & 4 & 1.4e-06 &  3.8 & 1.6e-06 &  4.2 & 2.0e-06 &  5.2 & $-0.06$ & $-0.15$ & $-0.10$\\
4.00 & 2600 & 8.2 & 6 & 3.8e-06 &  6.3 & 4.0e-06 &  7.5 & 4.1e-06 &  5.9 & $-0.02$ & $-0.03$ & $-0.01$\\
4.00 & 2600 & 8.5 & 2 & 1.4e-06 & 16.0 & 1.4e-06 & 16.2 & 1.2e-06 & 15.8 & $ 0.00$ & $-0.07$ & $-0.07$\\ 
4.00 & 2600 & 8.5 & 4 & 3.4e-06 & 15.9 & 3.8e-06 & 15.8 & 4.1e-06 & 17.5 & $-0.05$ & $-0.08$ & $-0.03$\\
4.00 & 2600 & 8.5 & 6 & 5.8e-06 & 16.5 & 6.0e-06 & 16.2 & 6.3e-06 & 15.8 & $-0.01$ & $-0.04$ & $-0.02$\\
4.00 & 2600 & 8.8 & 2 & 8.0e-06 & 30.3 & 7.6e-06 & 32.3 & 8.3e-06 & 31.2 & $ 0.02$ & $-0.02$ & $-0.04$\\
4.00 & 2600 & 8.8 & 4 & 8.9e-06 & 29.5 & 1.0e-05 & 28.0 & 8.4e-06 & 29.7 & $-0.05$ & $ 0.03$ & $ 0.08$\\
4.00 & 2600 & 8.8 & 6 & 1.5e-05 & 25.7 & 1.6e-05 & 27.1 & 1.6e-05 & 25.8 & $-0.03$ & $-0.03$ & $ 0.00$\\
4.00 & 2800 & 8.2 & 6 & 1.4e-06 &  0.9  & 1.8e-06 &  2.9  & 2.2e-06 &  3.3  & $-0.11$ & $-0.20$ & $-0.09$\\
4.00 & 2800 & 8.5 & 4 & 2.2e-06 & 21.2 & 2.4e-06 & 19.4 & 2.5e-06 & 19.6 & $-0.04$ & $-0.06$ & $-0.02$\\
4.00 & 2800 & 8.5 & 6 & 3.2e-06 &  8.1 & 4.3e-06 & 10.1 & 4.6e-06 & 10.5 & $-0.13$ & $-0.16$ & $-0.03$\\
4.00 & 2800 & 8.8 & 2 & 2.4e-06 & 40.3 & 3.4e-06 & 33.4 & 3.5e-06 & 34.0 & $-0.15$ & $-0.16$ & $-0.01$\\
4.00 & 2800 & 8.8 & 4 & 8.8e-06 & 26.0 & 8.9e-06 & 26.3 & 1.1e-05 & 25.5 & $-0.00$ & $-0.10$ & $-0.09$\\
4.00 & 2800 & 8.8 & 6 & 8.3e-06 & 26.2 & 9.4e-06 & 26.0 & 1.1e-05 & 24.2 & $-0.05$ & $-0.12$ & $-0.07$\\
4.00 & 3000 & 8.2 & 6 & 9.4e-07 &  1.7  &              &         &              &         & 	 	  &               &      \\
4.00 & 3000 & 8.5 & 4 & 1.1e-06 & 12.3 & 1.6e-06 & 19.9 &              &         & $-0.16$ &               &      \\
4.00 & 3000 & 8.5 & 6 & 2.0e-06 & 16.7 & 2.0e-06 & 13.3 & 3.0e-06 & 14.3 & $ 0.00$ & $-0.18$ & $-0.18$\\
4.00 & 3000 & 8.8 & 4 & 4.4e-06 & 27.9 & 6.1e-06 & 23.7 & 6.5e-06 & 24.0 & $-0.14$ & $-0.17$ & $-0.03$\\
4.00 & 3000 & 8.8 & 6 & 6.0e-06 & 19.2 & 8.5e-06 & 23.2 & 8.9e-06 & 22.6 & $-0.15$ & $-0.17$ & $-0.02$\\
4.00 & 3200 & 8.8 & 4 & 6.8e-07 & 27.4 & 7.3e-07 & 26.2 & 5.2e-07 & 31.2 & $-0.03$ & $ 0.12$ & $ 0.15$\\
4.00 & 3200 & 8.8 & 6 & 3.7e-06 & 16.9 & 5.6e-06 & 12.9 & 5.8e-06 & 18.1 & $-0.18$ & $-0.20$ & $-0.02$\\
\hline 
Average &    &       &    &  	           &        &   	     &      	&             	&        & $-0.05$ & $-0.06$ & $-0.01$\\
\hline
\end{tabular}
\tablefoot{Columns 1-4 list input parameters (stellar luminosity, effective temperature, carbon excess and piston velocity). Columns 5-10 list the dynamical properties (mass-loss rate and wind velocity) of DARWIN models with three different metallicities (solar, LMC and SMC). Columns 11-13 list the differences in logarithmic mass-loss rates for models with the same input parameters but different metallicites. Note that all models have a current mass of one solar mass, $f_{\mathrm{L}}=2$ and pulsation periods given by the period-luminosity relation in \cite{feast1989}.}
\newline
\newline
\newline
\newline
\newline
\newline
\newline
\newline
\newline
\end{table*}

\section{Molecular influences on the radiation field}
\label{s_mol}

\begin{figure}
\centering
\includegraphics[width=0.9\linewidth]{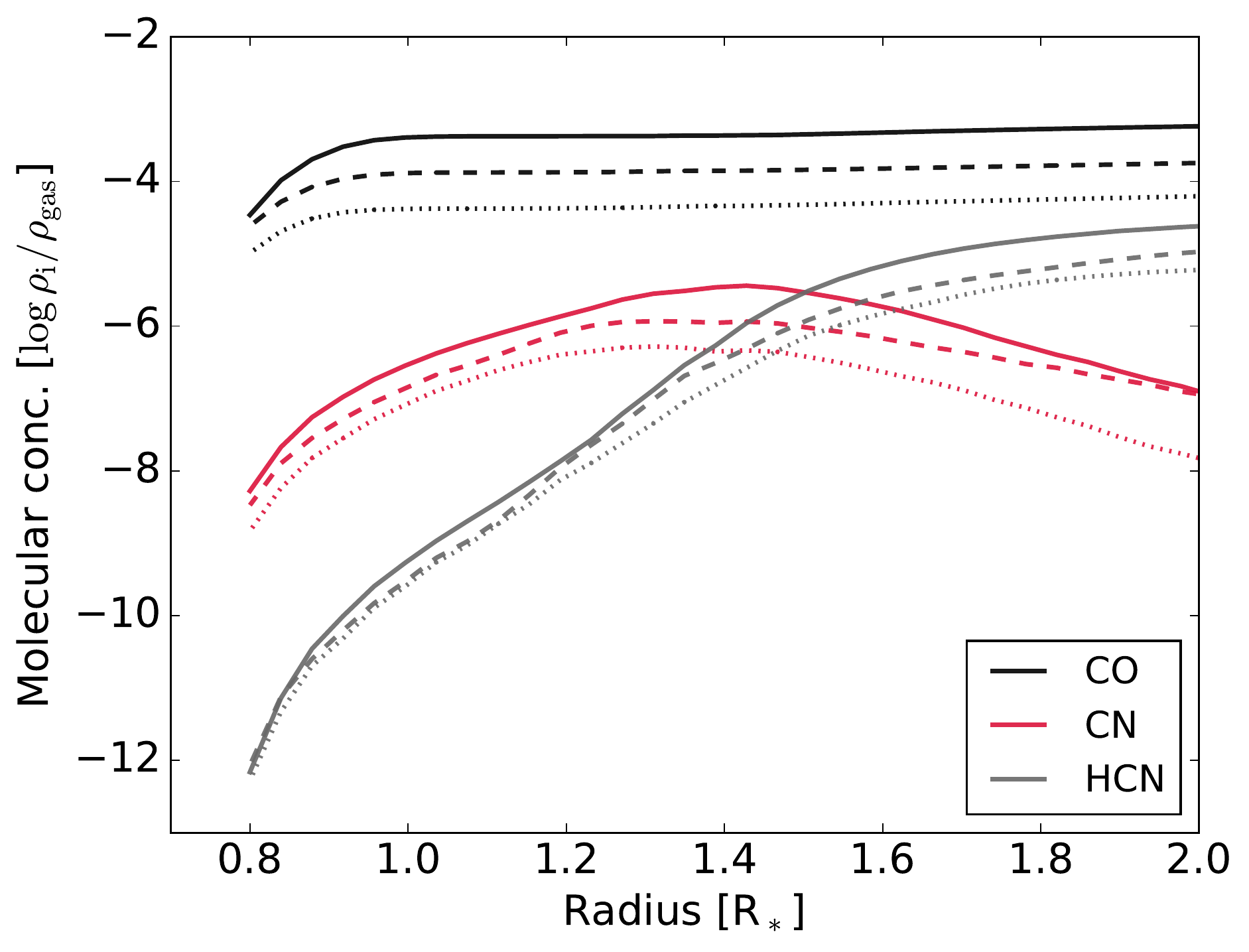} \\
\caption{Averaged molecular concentrations of CO, CN, and HCN as a function of radial distance for DARWIN models with input parameters $M_*=1\,\mathrm{M}_{\odot}$, $\log L_*/L_{\odot}=3.85$, $T_*=2800\,$K, $u_{\mathrm{p}}=4$\,km/s, and $\log\mathrm{(C-O)}+12=8.8$. The solid, dashed, and dotted lines show the molecular concentration from DARWIN models at solar, LMC, and SMC metallicities, respectively. }
\label{f_conc}
\end{figure}

The molecular abundances in carbon stars are affected by the chemical composition in the atmospheres and consequently by the metallicity. Figure~\ref{f_conc} shows the mean molecular concentration of CO, CN, and HCN for three DARWIN models with different metallicities, but otherwise the same input parameters ($M_*=1\,\mathrm{M}_{\odot}$, $\log L_*/L_{\odot}=3.85$, $T_*=2800\,$K, $u_{\mathrm{p}}=4$\,km/s, and $\log\mathrm{(C-O)}+12=8.8$). The concentration of CO is lower in the subsolar models since the formation of CO is limited by the amount of oxygen available. Similarly, the CN and HCN concentrations are lower since the abundances of these molecules are limited by the available nitrogen. This indicates that metal-poor carbon stars will have lower abundances of CO and higher C/O ratios than carbon stars at solar metallicity if the carbon excess is similar. The same reasoning also holds true for the CN and HCN abundances. For observational evidence of CN and HCN depletion in metal-poor carbon stars, see \cite{evans1980},\cite{cohen1981} and \cite{loon2008}.

Carbon dust typically condenses around two stellar radii, whereas the molecular species already form in the inner atmosphere. In order to investigate how the changes in molecular abundances with metallicity alter the radiation field, we calculate visual and near-IR photometric fluxes without including the contribution from dust opacities in the a posteriori radiative transfer. It should be noted that the atmospheric structures used to calculate these fluxes are based on hydrodynamic simulations where the effects of dust are included.

\begin{figure}
\centering
\includegraphics[width=0.95\linewidth]{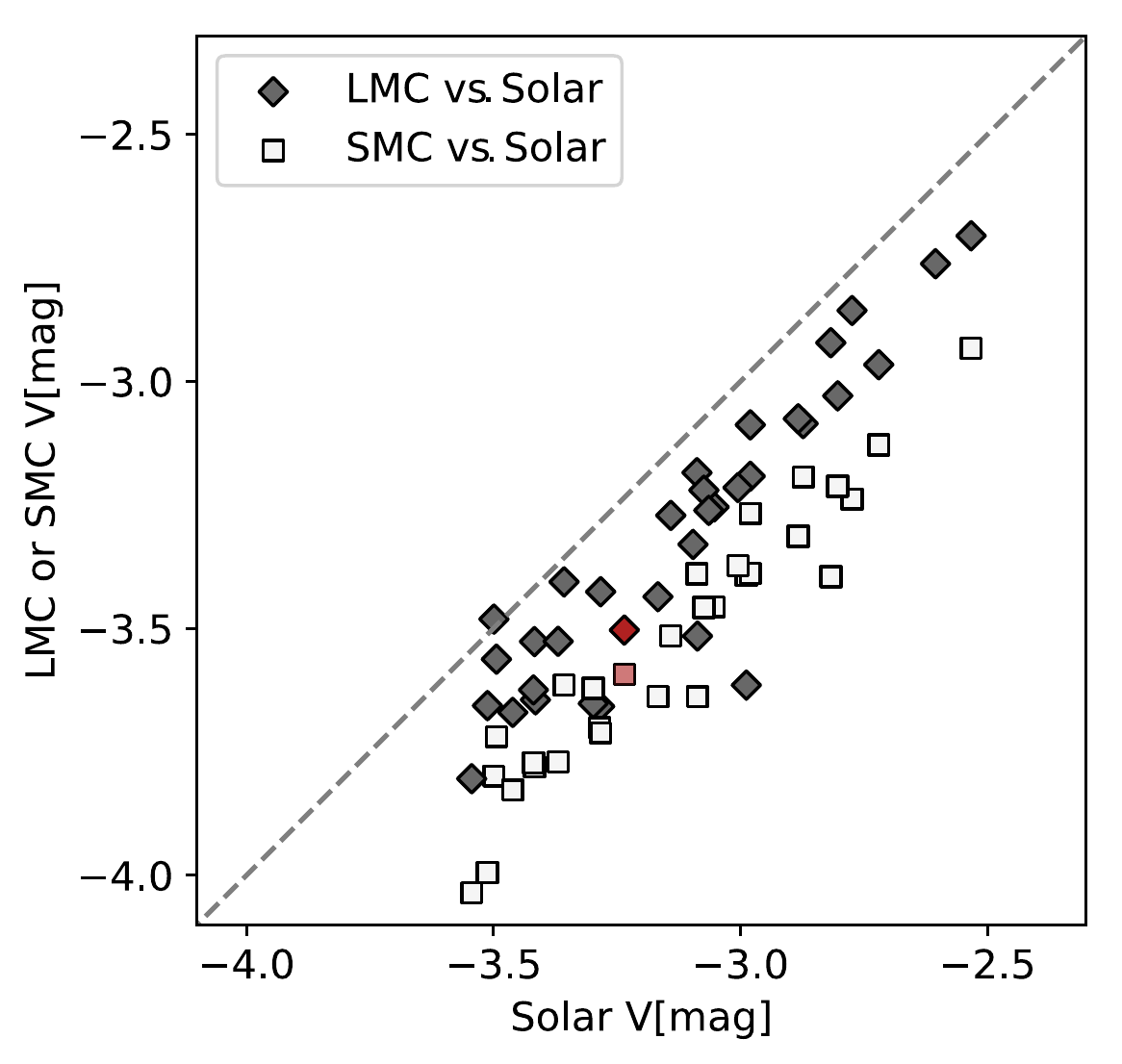}\\
\includegraphics[width=0.95\linewidth]{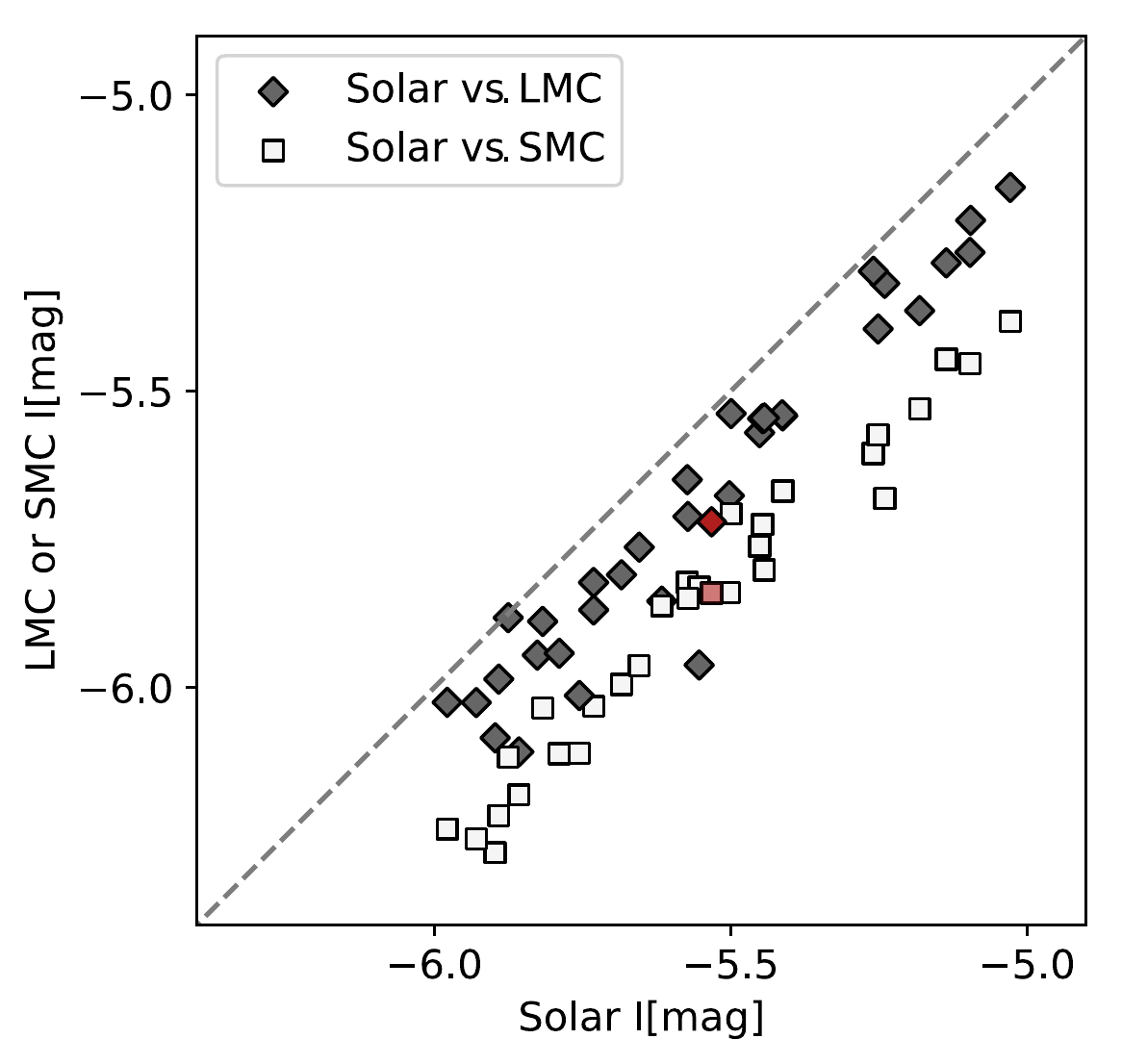}\\
\caption{Dust-free magnitudes in the $V$ (upper panel) and $I$ bands (lower panel) for models with the same input parameters but different metallicities plotted against each other. The magnitudes were calculated without including dust opacities in the a posteriori radiative transfer. The magnitudes of the DARWIN models used in Figs.~\ref{f_conc}-\ref{f_spec} are marked in red.}
\label{f_flux}
\end{figure}

\begin{figure}
\centering
\includegraphics[width=\linewidth]{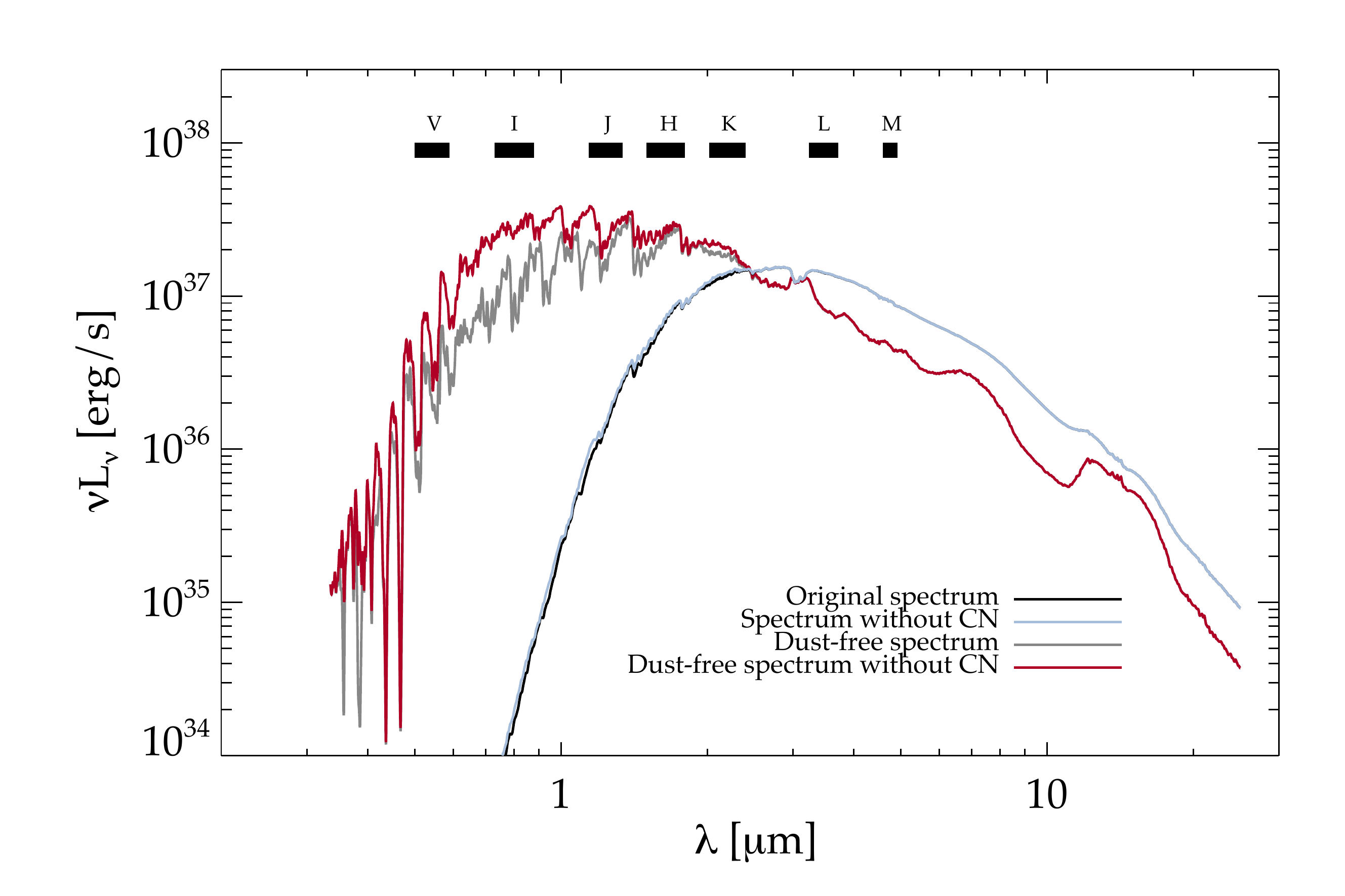} 
\caption{Spectra during luminosity minimum of a DARWIN model at LMC metallicity with input parameters $M_*=1\,\mathrm{M}_{\odot}$, $\log L_*/L_{\odot}=3.85$, $T_*=2800\,$K, $u_{\mathrm{p}}=4$\,km/s, and $\log\mathrm{(C-O)}+12=8.8$. The spectra including dust opacities are very similar.}
\label{f_spec}
\end{figure}

Figure~\ref{f_flux} shows the mean "dust-free" fluxes in the $V$ and $I$ bands from DARWIN models with the same input parameters but different metallicities, plotted against each other. If there is a one-to-one correspondence between models with the same stellar parameters but different metallicities, they will end up on the dashed line. As can be seen in Fig.~\ref{f_flux}, there is a systematic difference in the photometric fluxes at shorter wavelengths (below the $H$ band) depending on metallicity, corresponding to a stronger radiative flux at shorter wavelengths for models at subsolar metallicities. This increasingly bluer radiation field is mostly due to the decreased abundance of CN at subsolar metallicities, but atoms also influence the fluxes in the $V$ band (C$_2$ also strongly influences these photometric bands, but the abundance of C$_2$ is not affected much by a decrease in metallicity).  Figure~\ref{f_spec} shows spectra at minimum luminosity of a DARWIN model at LMC metallicity, calculated with and without dust opacities, and with and without CN opacities. The spectra with and without opacities for amorphous carbon nicely illustrate the redistribution of stellar radiation by the dust. The dust-free spectra with and without CN opacities show that changes in the molecular concentration of CN in the dust formation region will affect the photometric fluxes in the $J$, $I$, and $V$ bands. However, these changes will not be visible in the spectra due to the high optical depth of the dust shell.

This change in the radiation field provides a plausible explanation for the slightly shifted wind/no-wind boundary in parameter space at lower metallicities (see Fig.~\ref{f_windmaps}). The absorption cross section of amorphous carbon is approximately proportional to $\lambda^{-1}$, resulting in higher absorption cross sections at shorter wavelengths. The optical properties of amorphous carbon, together with the increasingly bluer radiation field at lower metallicities, make the newly condensed carbon grains more vulnerable to sublimation since the increased absorption of stellar light might cause the grains to heat up too much \citep[for an in depth discussion see][]{bladh2012}. 
If dust particles become thermally stable farther away from the star (where the density is lower) due to the harder stellar radiation it will be more difficult to drive a mass outflow.

\section{Implications for stellar evolution modeling}
\label{s_sem} 
\begin{figure}
\centering
\includegraphics[width=0.98\linewidth]{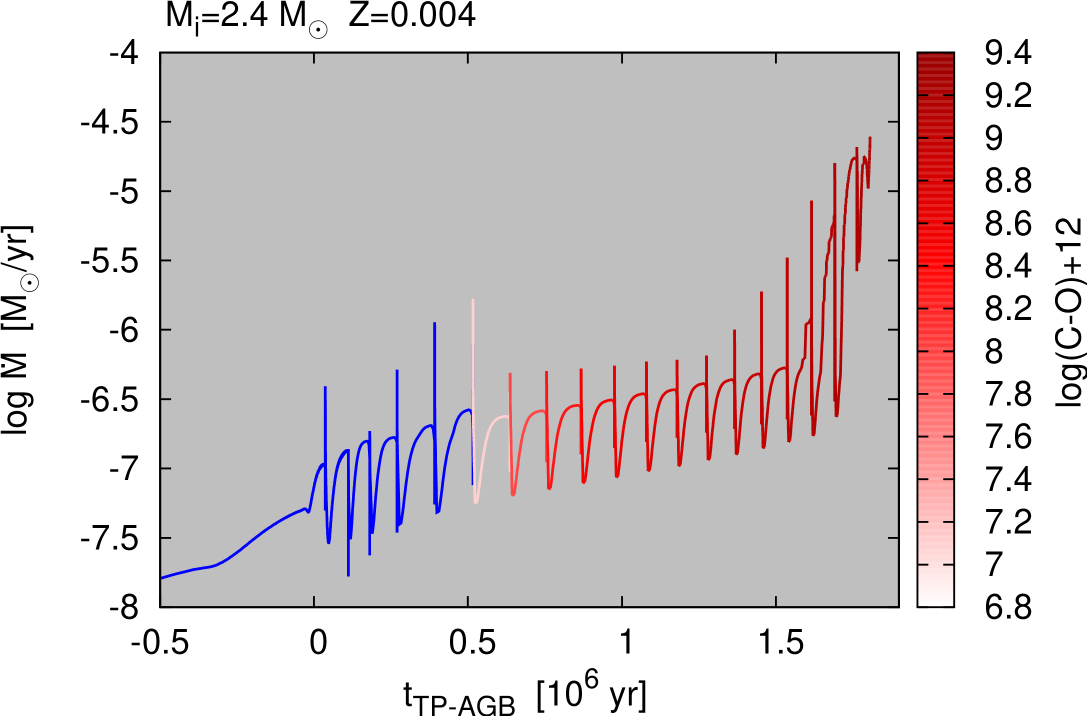}
\caption{Evolution of the  mass-loss rate during the final stages of the early AGB and the whole TP-AGB phase of a star with initial mass of 2.4 $M_{\odot}$ and metallicity $Z=0.004$, computed with the \texttt{COLIBRI} code \citep{marigo2013}. The zero of the time axis marks the occurrence of the first thermal pulse.
During the C-rich phases the track is color-coded according to the current carbon excess. See text for details.} 
\label{track}
\end{figure}

The results presented in Sect.~\ref{s_compdyn} show that mass-loss rates produced from wind models of carbon stars change on average 15\% if we lower the metallicity from solar to subsolar values, but keep the carbon excess constant. Given this result, it is clear that AGB stars can contribute to the interstellar dust production at lower metallicities if they manage to dredge up sufficient carbon from the stellar interior. Furthermore, this result shows that stellar evolution models can use the mass-loss rates calculated from DARWIN models at solar metallicity \citep[e.g the grid of][]{mattsson2010,eriksson14} when modeling the AGB phase at subsolar metallicities, as long as carbon excess is used as the critical abundance parameter instead of the C/O ratio. The effects of the increasingly harder radiation field with decreasing metallicity should be examined further before concluding that these results also hold for extreme metal-poor environments. Figure~\ref{track} provides an example of how mass-loss rates derived from DARWIN models can be used in stellar evolutionary calculations similar to those in \cite{marigo2013}. The selected stellar evolution model has an initial metallicity $Z=0.004$, which corresponds to [Fe/H]= -0.6 using \cite{caffau2009} abundances. Mass loss during the early stages where C/O~$<1$ (marked in blue) is described with the mass-loss formula introduced by \citet[][with an efficiency parameter $\eta=0.02$]{bloecker95}, whereas the mass-loss rates during the later stages where C/O~$>1$ (marked in red) are obtained through interpolation in grids of dynamical models by \cite{mattsson2010} and \cite{eriksson14}. The carbon excess required to trigger a stellar wind varies with the current stellar parameters, as can be seen in Fig.~\ref{f_windmaps} (Sect.~\ref{s_windmaps}). However, below a certain threshold value there is not enough carbon (that is not bound in CO) available to form sufficient amounts of amorphous carbon grains to drive a wind. This threshold value is noticeable in Fig.~\ref{track} by the sudden increase in mass-loss rate as soon as the carbon excess overcomes $\log({\rm C-O})+12 \simeq$ 8.1. For values below this limit (but still with C/O~$>1$) the mass-loss rates are computed according to the model for cool chromospheres developed by \citet{CranmerSaar_11}. 

Figure~\ref{f_wachter} shows the mass-loss rates derived from DARWIN models at solar (top panel) and subsolar metallicities (middle and bottom panels) compared to the mass-loss formulas given in \cite{wachter2008}. These formulas are based on an older generation of dust-driven wind models that use gray radiative transfer and a constant value for the gas opacity, in contrast to the frequency-dependent radiative transfer and detailed gas and dust opacities used in DARWIN models (see Sect.~\ref{s_darwin}). These differences in the treatment of gas opacities and radiative transfer affect the density and temperature structures of the atmosphere. The low value chosen for the gas opacity in the models by \cite{wachter2008} results in much higher gas densities at a given temperature, which in turn translates to higher mass-loss rates. These models also have lower C/O ratios compared to the DARWIN models, which leads to slow wind velocities. It is clear from Fig.~\ref{f_wachter} that the mass-loss rates derived from the DARWIN models are systematically lower (a factor 3-6 lower on average, depending on metallicity) than the values obtained from the mass-loss formulas for the same stellar parameters. Using the formula by \cite{wachter2008} in stellar evolution models will therefore lead to an overestimation of the mass loss. 

\section{Conclusions and summary}
\label{s_concl} 
 We explore the metallicity dependence of mass loss by calculating DARWIN models of C-type AGB stars at different chemical abundances (solar, LMC, and SMC). The raw material for the wind-driving dust species in carbon stars (amorphous carbon) is the carbon that is not bound in CO molecules, i.e., carbon that instead can be found in molecular species such as C$_2$, C$_2$H, and C$_2$H$_2$. The most relevant quantity for dust formation, and consequently the mass loss, in these stars is therefore the carbon excess (C-O) rather than the C/O ratio. Since carbon may be dredged up during the thermal pulses as AGB stars evolve, we keep the carbon abundance as a free parameter and compare wind properties from models with the same carbon excess but different metallicities.

The dynamical output from these wind models shows that mass loss in carbon stars is facilitated by high luminosities, low effective temperatures, and high carbon excess, at both solar and subsolar metallicities. Similar combinations of effetive temperature, stellar luminosity, and carbon excess produce outflows, independent of metallicity, even though the parameter space for models producing a wind is slightly diminished at lower metallicities (see Fig.~\ref{f_windmaps}). This small change is probably due to an increasingly strong radiative flux at shorter wavelengths as the metallicity decreases, mostly caused by less extinction from the lower CN and atomic abundances. The amorphous carbon grains may heat up too much when interacting with the bluer radiation field, making it more difficult to drive an outflow. The effects of the increasingly hard radiation field with decreasing metallicity should be examined further before concluding that these results also hold for extreme metal-poor environments. We note that since the oxygen abundance is set by the overall metallicity, a metal-poor carbon star will have a lower abundance of CO and a higher C/O ratio than a carbon star at solar metallicity with the same carbon excess (see Table~\ref{t_opacity} and Fig.~\ref{f_conc}). Similarly, the CN and HCN abundances will be lower since the nitrogen abundance is lower in metal-poor carbon stars.

A closer examination of the wind properties from these models show that there is no large systematic trend in the mass-loss rates and wind velocities with respect to metallicity (see Fig.~\ref{f_dynall}), nor any systematic difference concerning the distribution of grain sizes or how much carbon is condensed into dust (see Fig.~\ref{f_grain}). For both solar and subsolar environments typical grain sizes are $0.1-0.5\,\mu$m, the degree of condensed carbon varies between 5\% and 40\% and the gas-to-dust ratios range between 500 and 10000. A comparison between the mass-loss rates produced by DARWIN models with the same input parameters, but different metallicity, shows that the average relative difference is approximately 15\%, and at most around 60\% (See Fig.~\ref{f_dyncomp} and Table~\ref{t_dynprop}).

\begin{figure}
\centering
\includegraphics[width=0.9\linewidth]{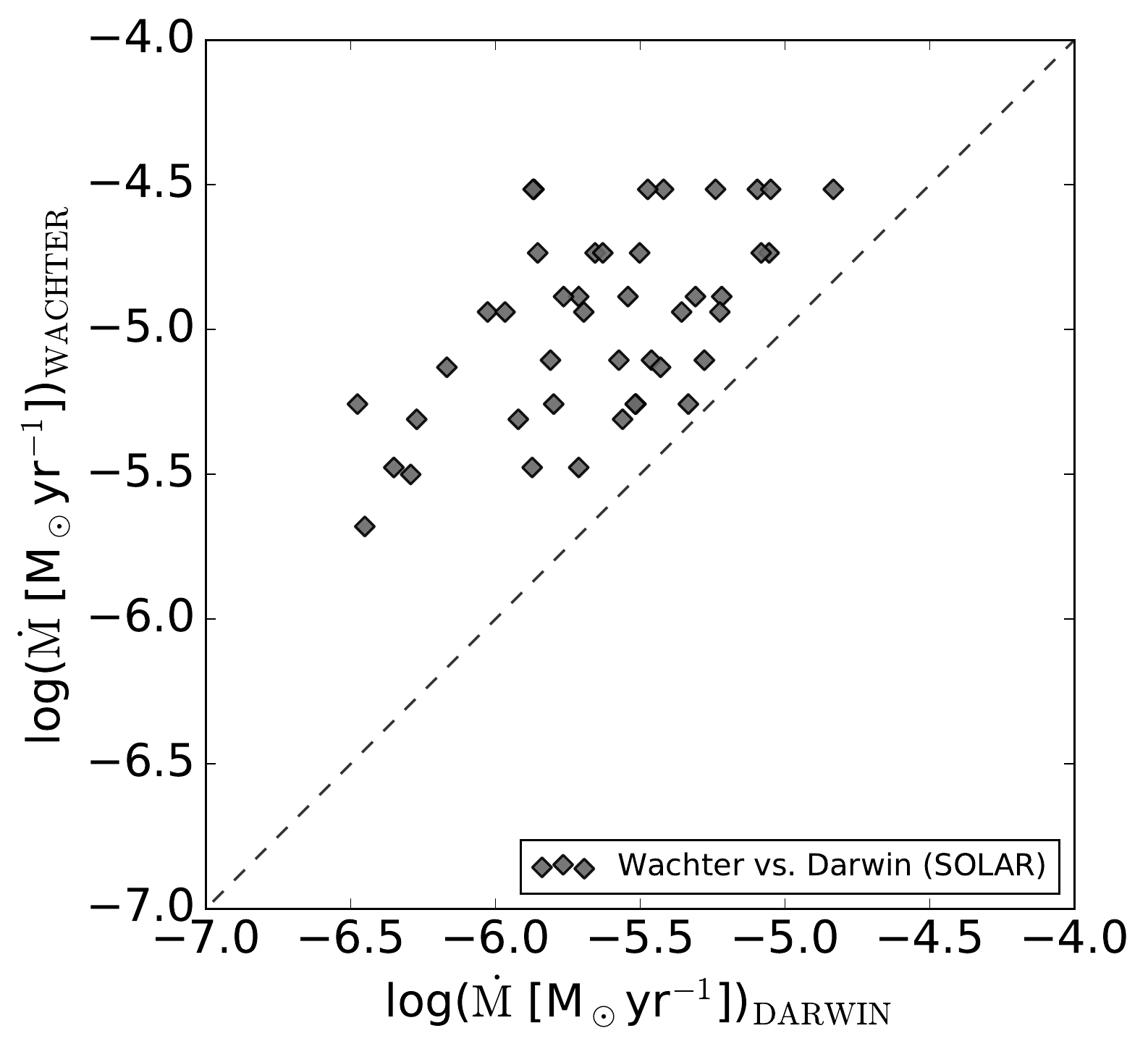}\\
\includegraphics[width=0.9\linewidth]{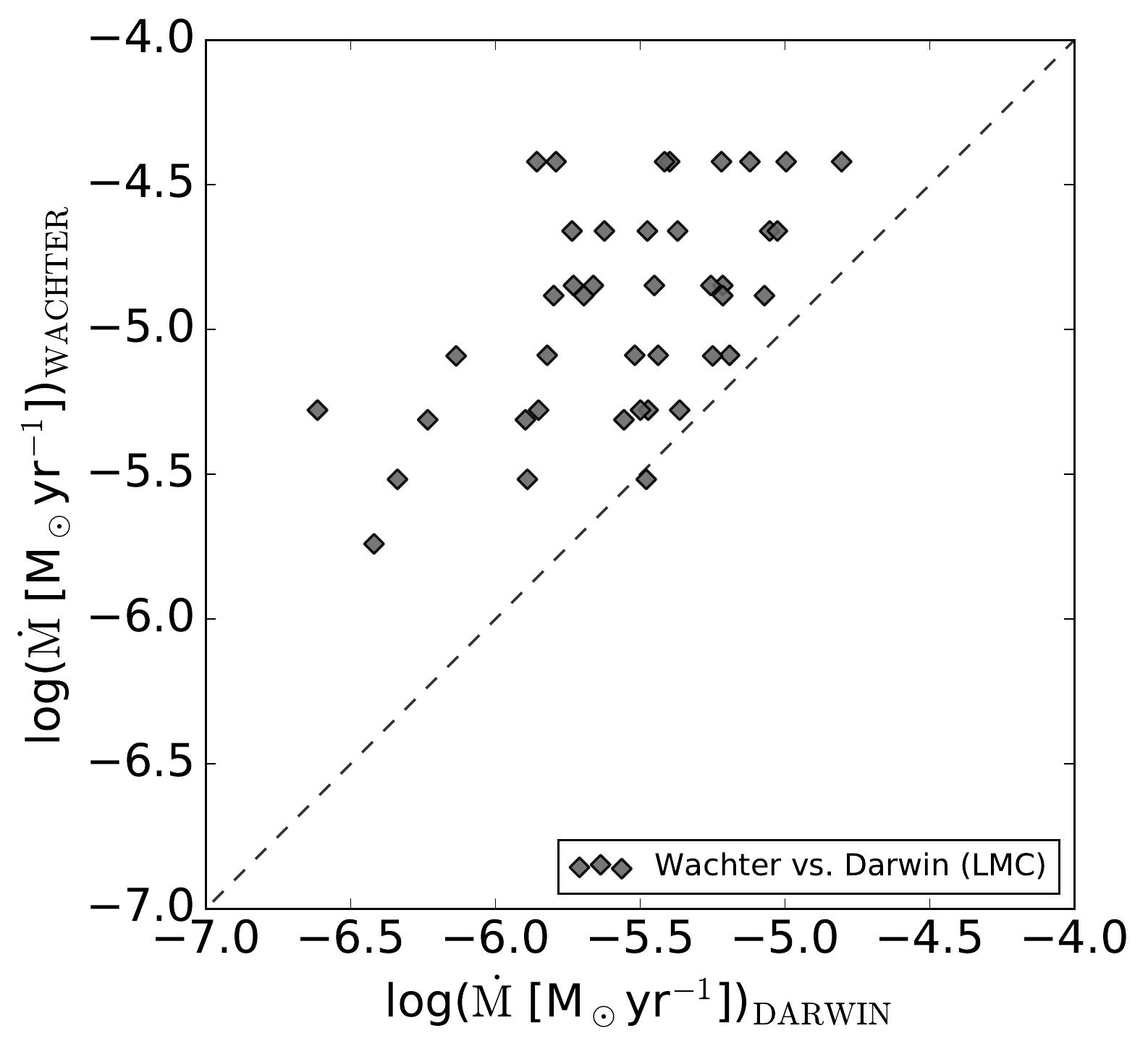}\\
\includegraphics[width=0.9\linewidth]{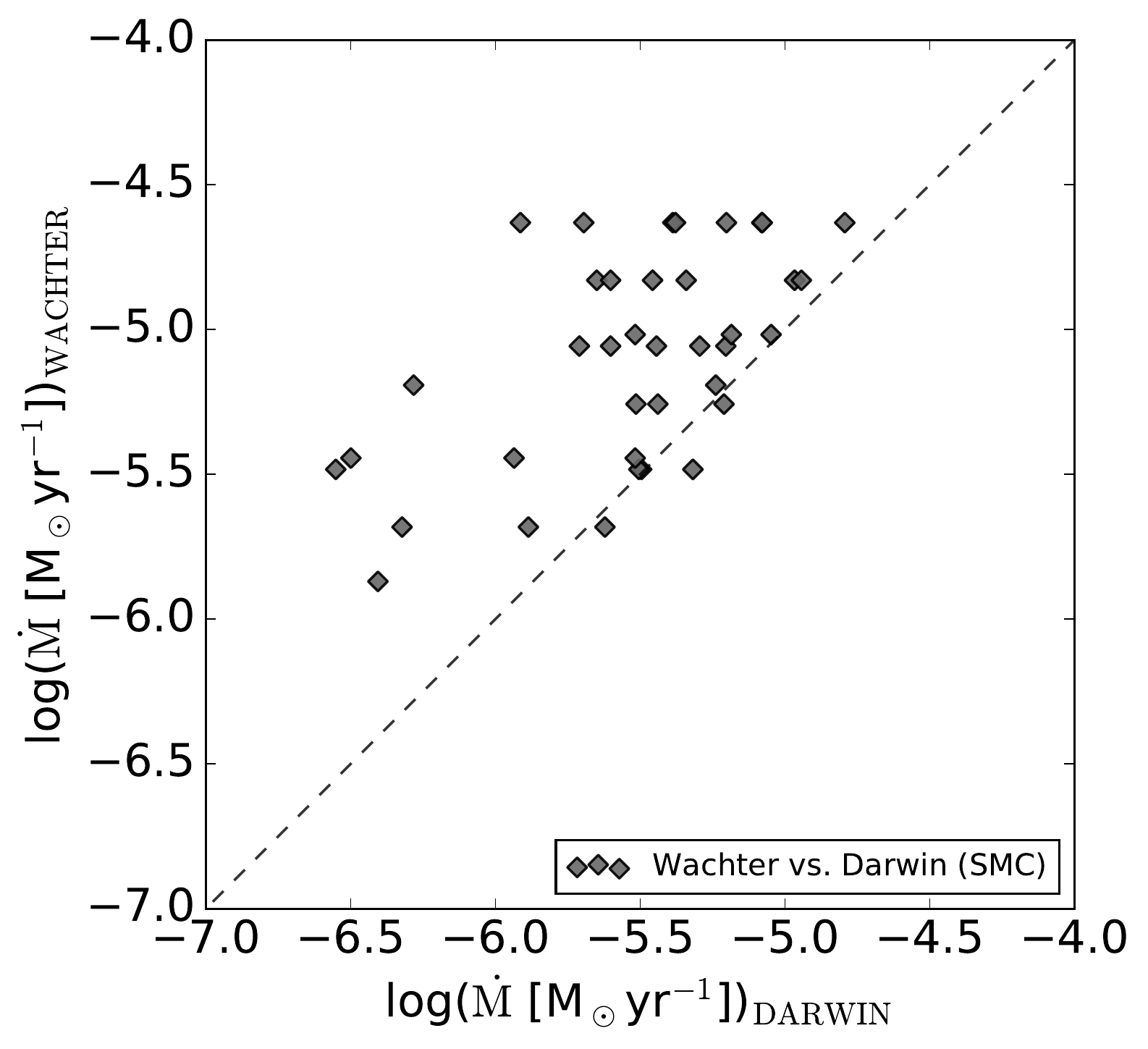}
\caption{Mass-loss rates derived from DARWIN models vs. mass-loss rates given by the mass-loss formulas in \cite{wachter2008} for the same stellar parameters. The upper, middle, and lower panels show mass-loss rates at solar ($[\mathrm{Fe/H}]=0$), LMC ($[\mathrm{Fe/H}]=-0.5$), and SMC ($[\mathrm{Fe/H}]=-1.0$) metallicity, respectively.}
\label{f_wachter}
\end{figure}

These results indicate that as long as carbon stars dredge up sufficient amounts of carbon from the stellar interior they can contribute to dust production in the ISM even in metal-poor environments, at least down to $[\mathrm{Fe/H}]=-1$ dex. This is confirmed by observations of dusty metal-poor carbon stars in the Local Group \citep[e.g.][]{sloan2012,boyer2017}. Given that the relative difference in mass-loss rates between DARWIN models with the same input parameters but different metallicities is on average 15\%, whereas the uncertainty in the observed mass-loss rates is estimated to reach as high as a factor of three \citep{Ramstedt2008}, stellar evolution models can use the mass-loss rates calculated from DARWIN models at solar metallicity \citep[e.g the grid of][]{mattsson2010,eriksson14} when modeling the AGB phase at subsolar metallicities as long as carbon excess is used as the critical abundance parameter instead of the C/O ratio. Furthermore, the mass-loss rates derived from DARWIN models are systematically lower (a factor of 3-6 lower on average, depending on metallicity, see Fig.~\ref{f_wachter}) than the values obtained from the mass-loss formulas presented by \cite{wachter2008} for the same stellar parameters. Using these formulas, based on an older generation of dust-driven wind models, will therefore lead to an overestimation of the mass loss in stellar evolution models.  

\begin{acknowledgements}
This work has been supported by the ERC Consolidator Grant funding scheme (project STARKEY, G.A. No. 615604) and the Schönberg donation. The computations were performed on resources provided by the Swedish National Infrastructure for Computing (SNIC) at UPPMAX.
\end{acknowledgements}


\bibliographystyle{aa}
\bibliography{biblio}

\end{document}